\documentclass[12pt]{article}
\usepackage{epsfig, bm, amsfonts, amssymb}

\RequirePackage{color}

\usepackage{graphicx}
\RequirePackage{color}

\definecolor{MyDarkGreen}{rgb}{0.02,0.60,0.06}

\usepackage{soul}

\date{December 2012}
\title{\bf Hyperscaling above the upper critical dimension}
\author{ 
{\it B. Berche$^{\,1}$,} {\it R. Kenna$^{\,2}$} and {\it J.-C. Walter$^{\,3,4}$}\\~\\
$^1$ Statistical Physics Group, Institut Jean Lamour,\\
Nancy Universit{\'{e}}, UMR CNRS 7198  \\
54506 Vand{\oe}uvre les Nancy Cedex, France
{}\\~\\
$^2$ 
Applied Mathematics Research Centre,\\
Coventry University,\\
Coventry, CV1 5FB, England
{}\\~\\
$^3$ Instituut voor Theoretische Fysica, \\
Katholieke Universiteit Leuven,\\
Celestijnenlaan 200D B-3001 Leuven, Belgium
{}\\~\\
$^4$ Instituut-Lorentz, Universiteit Leiden, \\
P.O. Box 9506, 2300 RA Leiden, The Netherlands
{}\\~\\}
\begin{document}
\maketitle
\vspace{-0.75cm}
                      {\Large
                      \begin{abstract}
Above the upper critical dimension, the breakdown of hyperscaling is associated with dangerous irrelevant variables in the renormalization group formalism at least for systems with periodic boundary conditions.
While these have been extensively studied, there have been only a few analyses of finite-size scaling with free boundary conditions.
The conventional expectation there is that, in contrast to periodic geometries, finite-size scaling is Gaussian, governed by a correlation length commensurate with the lattice extent.
Here, detailed numerical studies of the five-dimensional Ising model indicate that this expectation is unsupported,
both at the infinite-volume critical point and at the pseudocritical point where the finite-size susceptibility peaks.
Instead the evidence indicates that finite-size scaling at the pseudocritical point is similar to that in the periodic case.
An analytic explanation is offered which allows hyperscaling to be extended beyond the upper critical dimension.
%
                        \end{abstract} }
%
  \thispagestyle{empty}
%
%
  \newpage
%
                  \pagenumbering{arabic}

\section{Introduction}

It is well known that standard finite-size scaling (FSS) is valid below the upper critical dimension $d=d_c$ when hyperscaling holds and where the correlation length is comparable  to the  extent $L$ of a system exhibiting a continuous phase transition \cite{FSS}.
Above $d_c$, standard hyperscaling breaks down and  critical behaviour  is described by mean-field exponents \cite{Br82}. 
FSS was analyzed for the ($d \ge d_c=4$)-dimensional $\phi^4$ theory and Ising model with periodic boundary conditions (PBC) \cite{Bi85,BNPY,BZJ85,RuGu85,RuGa85,Binder87,RiNi94,Mon1996,PaRu96,LuBl97a,LuBl97b,BlLu97c,Luijtenthesis,LuBi99,BiLuquote,Turks,JoYo05,LuMa11} following  analytical studies by Br{\'{e}}zin \cite{Br82}.
The breakdown of  FSS and hyperscaling  is attributed to dangerous irrelevant variables  in the renormalization-group (RG) approach \cite{FiHa83}.
To repair FSS, Binder introduced the {\emph{thermodynamic length}}  which  scales as a power of $L$  above $d_c$ and as the correlation length 
below $d_c$   \cite{Bi85,Binder87,LuBi99}.
It is  perhaps surprising that, although the theoretical framework 
is well developed, especially with PBC's \cite{BDT,Review}, FSS above the upper critical dimension was summarized by Binder et al as 
``a rather disappointing state of affairs - although for the $\phi^4$ theory in $d = 5$ dimensions all exponents are known, including those of the corrections to scaling, and in principle very complete analytical calculations are possible, the existing theories clearly are not so good'' \cite{BiLuquote,Review}.

In contrast to the PBC case, there have  been  few studies of $d>d_c$ 
systems with free boundary conditions (FBC) \cite{RuGa85,LuMa11},  which are complicated by additional scaling fields associated with boundaries in the RG picture \cite{BDT,Bi83Diehl86}.
This situation was  described in Ref.\cite{JoYo05} as ``poorly understood''.
In a recent numerical study of the $d=5$ Ising model with FBC's at the infinite-volume transition point,
the question of the FSS behaviour of the magnetic susceptibility $\chi_L$ was addressed~\cite{LuMa11}. 
(The specific heat is less interesting  because the associated critical exponent vanishes
and the function does not diverge.)
A naive extension of ordinary FSS to beyond the upper critical dimension would deliver  leading Gaussian behaviour  $\chi_L \sim L^{2}$.
However, it is well established (and well understood) that $\chi_L \sim L^{5/2}$ for 5D Ising systems with PBC's,
the adjusted behaviour being traced back to an effect of dangerous irrelevant variables in the renormalization-group approach. 
The question of whether $\chi_L \sim L^{2}$ or $\chi_L \sim L^{5/2}$ in 5D was examined numerically in Ref.\cite{LuMa11} with the conclusion that  the former  prevails at the infinite-volume critical point
in the FBC case.

Here we point out that, for the FBC lattice sizes used in Ref.\cite{LuMa11}, the bulk of sites are on the surface, so that the system is not genuinely five-dimensional. 
Therefore, the resulting  conclusion that $\chi_L \sim L^2$ is  not a 5D one.
For this reason, we re-examine FSS for the 5D Ising model.
Besides looking at the critical temperature (which was the  focus in Ref.\cite{LuMa11}), we also examine the pseudocritical temperature, given by the location of the susceptibility maximum.
In order to probe the five-dimensionality of the structure, 
we remove contributions close to the lattice boundary. 
A similar tactic of removing the contributions of boundary spins has been used to examine the Ising model on a negative-curvature substrate \cite{pseudosphere}.
Moreover, besides the susceptibility, we investigate the magnetization and the Lee-Yang zeros. 
Our numerical results indicate that, in contradistinction to the results of Ref.\cite{LuMa11} at criticality, the FBC lattice exhibits the {\emph{same}} FSS as the PBC one at pseudocriticality, namely $\chi_L \sim L^{5/2}$.
The FSS for the magnetization and Lee-Yang zeros similarly match.
These numerical observations indicate that the conventional  FSS paradigm for the susceptibility of FBC lattices above $d_c$  is {\emph{unsupported}} \cite{RuGa85,LuMa11, Watson,Gunton}.
We offer an analytical explanation for this phenomenon through a Lee-Yang analysis and suggest that hyperscaling can, in fact, be extended to beyond the upper critical dimension in a universal manner.

\section{Hyperscaling and finite-size scaling}

With the reduced temperature $t$ denoting  distance from the critical point, the leading scaling for the specific heat, spontaneous magnetization,  susceptibility and correlation length are
$c_\infty(t) \sim  |t|^{-\alpha}$,
$m_\infty(t) \sim  |t|^{\beta}$,
$\chi_\infty(t) \sim  |t|^{-\gamma}$ and
$ \xi_\infty(t) \sim  |t|^{-\nu} $,
respectively. 
Here  the subscript indicates the system size.
At $t=0$,   magnetization in field scales as $ m_\infty(h) \sim h^{1/\delta}$.
The   critical exponents are related through  scaling relations, of which hyperscaling
\begin{equation}
\nu d  =  2-\alpha  \, \quad {\mbox{for}} \quad d \le d_c
\label{hyperscaling}  
\end{equation}
is conspicuous in that it involves the dimensionality $d$ and 
fails above the   critical dimension, which is $d_c=4$ for the Ising model.
There, Landau's  mean field exponents are
$\alpha =  0$, $\beta  = 1/2$, $\gamma  = 1$, 
$\delta  = 3$, $\nu  = 1/2$, $\eta = 0$ (the last of these is the anomalous dimension).

FSS was formulated by Fisher \cite{FSS} and  derived by Br{\'{e}}zin  in the absence of a singularity at the RG fixed point \cite{Br82} (see Refs.~\cite{BDT,Ba83,Pr90} for reviews).
Privman and Fisher \cite{PrFi84} showed that the singular part of the finite-size free energy below the upper critical dimension is 
\begin{equation}
 f_L(t,h) = b^{-d} f_{L/b}
 \left({
  tb^{y_t},hb^{y_h}
 }\right),
\label{woDIV}
\end{equation}
a result confirmed  for the spherical model in Ref.~\cite{SiPa85}. 
Field-theoretic, $\epsilon$-expansion calculations of finite-size scaling functions followed in 
Refs.\cite{BZJ85,RuGu85}. 
At $h=0$, Eq.(\ref{woDIV}) can be written 
$ f_L(t,0) = L^{-d} f_{1} \left({  tL^{y_t},0 }\right) = L^{-d} f_{1} \left({(L/\xi_\infty )^{1/\nu},0 }\right) $.
Thus standard FSS is controlled by the first argument on the right-hand side, namely 
the ratio $x=L/\xi_\infty(t)$. When $x \ll 1$ the system is in the scaling regime and the function
$f_1(x^{1/\nu})$ is universal, behaving to leading order with a power law $f_1(x^{1/\nu})\sim x^d$
in such a way that $f_L(t)\sim t^{d\nu}$,
while outside this regime the physical quantities depend on both $t$ and $L$ in a non-universal
and non-trivial manner. 
The  correlation length is
\begin{equation}
\xi_L(t,h) = b \Xi_{L/b} \left({tb^{y_t},h b^{y_h}}\right).
\label{BNPY200}
\end{equation}
Setting $b=L$ and $h=0$, the last equation gives $\xi_\infty (t,0) \sim t^{-1/{y_t}}$ when $L \rightarrow \infty$ and 
identifies $\nu = 1/y_t$.
At $t=0$, it gives $\xi_L \sim L $.

Above the upper critical dimension, the critical behaviour is controlled by the Gaussian fixed point  where \cite{Ma}
\begin{equation}
 y_t = 2, \quad y_h = 1+\frac{d}{2}.
\label{scdim}
\end{equation}
 Naively differentiating Eq.(\ref{woDIV}) delivers thermodynamic-scaling functions different to those from Landau mean field theory \cite{Landau}.
Taking irrelevant scaling fields  into account, however, gives a more complete version of Eq.(\ref{woDIV}) \cite{FiHa83},
\begin{equation}
f_L(t,h,u) = b^{-d} f_{L/b} \left({tb^{y_t},hL^{y_h},uL^{y_u}}\right),
\label{BL4.2} 
\end{equation}
with
\begin{equation}
  y_u = 4-d.
\label{scdim+}
\end{equation}
Here $u$ is the field associated with the coefficient of the $\phi^4$ term in the Ginzburg-Landau-Wilson expansion.
Below $d_c$,  irrelevant fields  contribute  to Wegner corrections \cite{Wegner72}, while at and above $d_c$, 
$u$ becomes marginal and then  dangerously irrelevant and leads  to the breakdown in FSS \cite{Br82}.

Motivated by Fisher's considerations in the thermodynamic limit \cite{FiHa83}, the behaviour of Eq.(\ref{BL4.2}) in the $u \rightarrow 0$ limit above $d_c$ was addressed in Ref.~\cite{BNPY}. The approach of
Fisher with dangerous irrelevant variables finds its origin in the mean-field (MF) solutions of
Landau theory governed by the free energy 
\begin{equation}
F[m(x)]=\int d^d x\left({\frac 12 r_0m^2+\frac 14um^4-hm+\frac 12|\vec\nabla m|^2}\right).
\end{equation}
(We choose to use $m(x)$ instead of the more conventional
$\phi(x)$ for the order parameter to be consistent with later notation.)
Minimizing the free energy leads to the thermodynamic quantities.
For example the  order parameter at low temperature in zero external field is
$m_0(t^-)=(-r_0/u)^{1/2}$. It is clear that the limit $u\to 0$ is singular, and this is
taken into account with the scaling assumption
\begin{equation}
m_L(t,0,u) = b^{-d}(ub^{y_u})^\kappa m_{L/b} \left(tb^{y_t}\right),
\label{eqn-m-div} 
\end{equation}
with $\kappa=-1/2$ from the MF solution. The exponent $\beta$ of the magnetization in zero field 
then follows 
\begin{equation}
\beta=\frac{d-y_h-\kappa y_u}{y_t}
\end{equation}
 instead of the usual expression  $({d-y_h})/{y_t}$. 
In a similar manner, the other standard exponents are given in terms of
the dimension $y_u$ of the dangerous irrelevant variable, 
\begin{equation}
\gamma=\frac{2y_h-d}{y_t},\ 
\delta=\frac{y_h}{d-y_h-\lambda y_u},\ \alpha=\frac{2y_t-d+\mu y_u}{y_t}
\end{equation}
with $\lambda=-1/3$ and $\mu = -1$ (following from MF expressions $m_0(h)=(h/u)^{1/3}$ and
$C(t^-)=(r_0/t)^2/u$).
The role of the dangerous variable can be incorporated in the definition of 
new scaling dimensions for the relevant fields, $t$ and $h$, as~\cite{BNPY,BZJ85}
\begin{equation}y_t^*
=y_t+p_2 y_u=\frac{d}{2}\label{scytstar}\end{equation} and
\begin{equation}y_h^*
=y_h+p_3 y_u=\frac{3d}{4}\label{scyhstar}\end{equation} with
$p_2=\mu y_t/(d-\mu y_u)=-\frac 12$ and
$p_3=\kappa(2y_h-d)/(d-2\kappa y_u)=-\frac 14$ in such a way that the usual definitions of exponents become valid and
Landau exponents are recovered \cite{BNPY,BZJ85},
$\beta=({d-y_h^*})/{y_t^*}= 1/2$, $\gamma=({2y_h^*-d})/{y_t^*}=1$,
$\delta={y_h^*}/({d-y_h^*})=3$ and $\alpha=({2y_t^*-d})/{y_t^*}=0$.

Taking these elements into account, the behaviour of Eq.(\ref{BL4.2}) in the $u \rightarrow 0$ was addressed in Ref.~\cite{BNPY} and
\begin{equation}
f_L(t,h,u) = b^{-d} f_{L/b} \left({tb^{y_t^*},hb^{y_h^*}}\right)
= L^{-d} f_{1} \left({tL^{y_t^*},hL^{y_h^*}}\right).
\label{BNPY}
\end{equation}
Similar considerations for the correlation length deliver 
\begin{equation}
\xi_L(t,h,u) = L^q \Xi \left({tL^{y_t^{*}},h L^{y_h^{*}}}\right).
\label{BNPY2}
\end{equation} 
The assumption $q=1$ was driven by the belief at the time that ``the correlation length $\xi_L$ is bounded by $L$'' even for PBC's \cite{BNPY}. 
In this case a second length scale would be needed to modify FSS \cite{Bi85,Binder87}.

The first argument on the right-hand side of  Eq.(\ref{BNPY}) or Eq.(\ref{BNPY2}) may be written $(\ell_\infty(t)/L)^{y_t^*}$ where 
$\ell_\infty(t) \sim t^{-1/y_t^*}$ and was dubbed the {\emph{thermodynamic length}} by Binder \cite{Bi85}. 
The finite-size {\emph{coherence length}} $\ell_L$ of Brankov, Danchev and Tonchev is the FSS counterpart of the thermodynamic length and therefore scales as the system extent $L$ \cite{BDT}. They also introduced a {\emph{characteristic length}} $\lambda_L(t)$, which is the FSS counterpart of the infinite-volume correlation length. 

For the 5D PBC model, a direct, explicit, numerical calculation of the FSS of the correlation length 
showed $\xi_L \sim L^{5/4}$ rather than $L$ as in standard FSS theory \cite{JoYo05}.
It is by now well established that the replacement of the scaling variable $L/\xi_\infty(t)$ of standard FSS by $L^{d/d_c}/\xi_\infty(t)$ of modified FSS is correct for the susceptibility, magnetization and pseudocritical point in periodic Ising models in four \cite{Luijtenthesis,JoYo05,KeLa91}, five \cite{RiNi94,PaRu96,Luijtenthesis,JoYo05,AkEr99}, six \cite{AkEr00,MeEr04,MeBa05}, seven \cite{AkEr01} and eight \cite{MeDu06} dimensions, 
giving strong support to the assertion that $q=d/d_c$.

To summarize, in all dimensions (including at the critical one), FSS is obtained by fixing  $\xi_L/\xi_\infty(t)$,
or, equivalently by making the replacement \cite{FSS,KeLa91}
\begin{equation}
 \xi_\infty(t) \rightarrow \xi_L \quad {\mbox{or}} \quad t \rightarrow \xi_L^{-{1}/{\nu}}
 \label{referee}
\end{equation}
in the scaling formulae for the various thermodynamic functions.
Below $d=d_c$, one has that the finite-size correlation length  scales as $\xi_L \sim L$ and the 
replacement (\ref{referee}) generates the correct FSS at both the pseudocritical and critical points.
Above $d=d_c$, one has instead that  
\begin{equation}
 {\xi}_L \sim L^q , \quad {\mbox{where}} \quad q=\frac{d}{d_c},
\label{q}
\end{equation}
which arises through dangerous irrelevant variables, at least for PBC's \cite{Br82,Bi85,BNPY,BZJ85,RuGu85,RuGa85,Binder87,RiNi94,Mon1996,PaRu96,LuBl97a,LuBl97b,BlLu97c,Luijtenthesis,LuBi99,BiLuquote}. 
We shall see that for $d>d_c$ the prescription (\ref{referee}) is also valid at pseudocriticality and at criticality in the PBC case and that it  holds at pseudocriticality, but not at criticality, in the case of FBC's.
Let us emphasize the fact that Eq.~(\ref{BNPY2}), although  written 
explicitly in Ref.~\cite{BNPY}, was immediately discarded in favour of $q=1$,
leaving room for inconsistencies in the standard dangerous irrelevant variable
scenario.

Eq.(\ref{q}) may be understood heuristically by demanding that the  volume associated with $\xi_L$ (in $d_c$ dimensions) correspond to the actual volume of the system,
\begin{equation}
 \xi_L^{d_c} \sim L^d
 .
 \label{volume}
\end{equation} 
The replacement $\xi_\infty(t) \sim  {\xi}_L$  yields $ |t| \sim L^{-{q}/{\nu}} $,
so that the susceptibility scales as
\begin{equation}
 \chi_L \sim L^{\frac{\gamma q}{\nu}} = L^{2q} 
 .
\label{FSSchi}
\end{equation}
Similarly, the FSS for the magnetization is 
\begin{equation}
 m_L \sim L^{-\frac{\beta q}{\nu}} = L^{-q} 
 .
\label{FSSm}
\end{equation}
Evidence from a variety of studies supports Eqs.(\ref{FSSchi}) and (\ref{FSSm}) with $q=d/d_c$ for PBC's 
\cite{Br82,Bi85,BNPY,BZJ85,RuGa85,Binder87,RiNi94,Mon1996,PaRu96,LuBl97a,LuBl97b,BlLu97c,Luijtenthesis,LuBi99,BiLuquote,Turks,JoYo05,LuMa11}, but the prevailing picture is that $q=1$ for FBC's \cite{RuGa85,LuMa11,Watson,Gunton}.
To distinguish between the two scenarios above the upper critical dimension, we refer here to the $q=1$ set-up as ``Gaussian'' and the
$q=d/d_c$ variant as ``$Q$'' scaling or $Q$-FSS.
  
We are interested in studying various functions at the infinite-volume critical point  $(t,h)=(0,0)$.
In the absence of field, we are also interested in the pseudocritical point defined, for example, by the location of the susceptibility peak $t=t_L$. 
The shift exponent $\lambda$ is associated with scaling of the pseudocritical point $t_L$ and usually (but not always) 
coincides with the  inverse of the correlation length critical exponent $1/\nu$ \cite{JaKe02}.
It characterises how the peak in the susceptibility moves in response to changing lattice size.
This peak is a smoothening out of the  divergence present in the thermodynamic limit and is also associated with the rounding exponent $\theta$.
The rounding may be  defined as the width at half-height of the susceptibility, so that \cite{Schladming}
\begin{equation}
 t_L \equiv \frac{|T_L-T_c|}{T_c} \sim L^{-\lambda}, \quad \quad \quad \Delta T_{\rm{rounding}} \sim L^{-\theta},
 \label{shiftround1}
\end{equation}
where $T_c$ is the critical value of the temperature $T$.
The naive  predictions for the rounding and shifting exponents coming from Gaussian FSS are 
\begin{equation}
 \lambda = \theta = \frac{1}{\nu} ,
\label{shiftround2}
\end{equation}
while the corresponding $Q$-FSS behaviour may be expected to be
\begin{equation}
 \lambda = \theta = \frac{q}{\nu} .
\label{shiftround3}
\end{equation}
It is also possible to study the shifting in terms of field from $H=0$ 
to $H=H_L$ (and associated rounding  if desired). 
We can  define a shift exponent associated with the effective critical 
field, $H_L \sim L^{-\lambda_H}$. 
One expects the usual relation $\lambda_H = \beta \delta / \nu$ to be modified to $\lambda_H = q\beta \delta / \nu$ 
with $q=d/4$ above $d_c=4$ dimensions and with PBC's. Again, the usual expectation is that $q=1$ for FBC's,
similar to below the upper critical dimension.

\section{Numerics}

To investigate the nature of the exponent $q$, we performed Monte Carlo simulations of the Ising model using the single-cluster Wolff algorithm \cite{Wo89} in combination with histogram reweighting \cite{FeSw88,FeSw89}. 
This approach is suitable to obtain reasonable correlation times for the lattice sizes considered, working close to the critical temperature. 
We simulated the $d=5$ Ising model on simple hypercubic lattices with sizes from $L=11$ to $L=51$.
The value of the critical temperature is known from Ref.~\cite{Review} to be $\beta_c=1/T_c=0.113\,915\,5(5)$ 
and our own simulations are in excellent agreement with this. 
To obtain an estimate  of a quantity at the pseudo-critical temperature, we first made  a rough estimation of this point. Then, the estimate is obtained by a longer simulation in  close vicinity to the first one. 
Thus every simulation has been performed, within error bars, at the critical point $T_c$ or at the pseudocritical point $T_L$, so that reweighting is not required except to estimate the rounding. 
For each lattice size, we  calculated the correlation time $\tau$ at the temperature of simulation. The samplings are spaced by $2\tau$ iterations of the Wolff algorithm, so that the configurations can be considered independent. 
For the rounding, the full error is obtained using standard techniques for histogram reweighting \cite{reweightingpaper}. 
In addition,  we  checked the overlap of histogram between the two values of temperature $T_c$ and $T_L$ from the 2 independent simulations.

Denoting the Ising spin at site $i$ of the lattice by $S_i$ the correlation length can be numerically calculated using the definition  \cite{Kim1993,CaGa01} 
(which holds only for PBC's and is the same as that used in Ref.~\cite{JoYo05})  
\begin{equation}
\xi_L=\frac{1}{k_{\rm{min}}}\sqrt{\tilde{G}(0)/\tilde{G}(k_{\rm{min}})-1}\,,
\label{SMxiL}
\end{equation}
where $k_{\rm{min}}=2\pi/L$ is the smallest wave vector of the periodic lattice and 
$\tilde{G}(k)$ is the Fourier transform of the spin-spin correlation function
\begin{equation}
\tilde{G}({\vec{k}})=\frac{1}{L^d}\sum_{i,j}
 \langle{S_iS_j}\rangle
 \exp{(i\vec k\cdot(\vec r_i-\vec r_j))},
\end{equation}
in which ${\vec{r}}_i$ marks the position of the $i^{\rm{th}}$
lattice site.
This  is calculated along the main axis of the lattice,
${\displaystyle{
G(i)=\langle{S_0S_i}\rangle-\langle{S_0}\rangle^2
}}$,
where $i$ runs between $0$ and $L/2$ in the PBC case. 
With ${\displaystyle{M=\sum_{i}{S_i}}}$, 
our numerical definitions of the magnetization and susceptibility are
\begin{equation}
 m_L(t) = \frac{1}{L^d}   \langle{|M|}\rangle\,, 
 \quad \quad \quad
 \chi_L(t) = \frac{1}{L^d}
  {
   \left({          \langle{M^2}\rangle -
          \langle{|M|}\rangle^2
         }
   \right)}\,,
\label{chinumeric}
\end{equation}
respectively.
Finally, it is useful to introduce the Binder cumulant, which is related to the kurtosis in statistics and the renormalized coupling constant \cite{Schladming},
\begin{equation}
 B =  \frac{\langle{M^4}\rangle}{\langle{M^2}\rangle^2}.
 \label{Bindercumul}
\end{equation}
For $\xi_L \gg L$, this varies only weakly with temperature, staying close to a universal but non-trivial value.

\subsection{Periodic boundary conditions}

\begin{figure}
  \begin{center}
  \includegraphics[width=0.8\columnwidth]{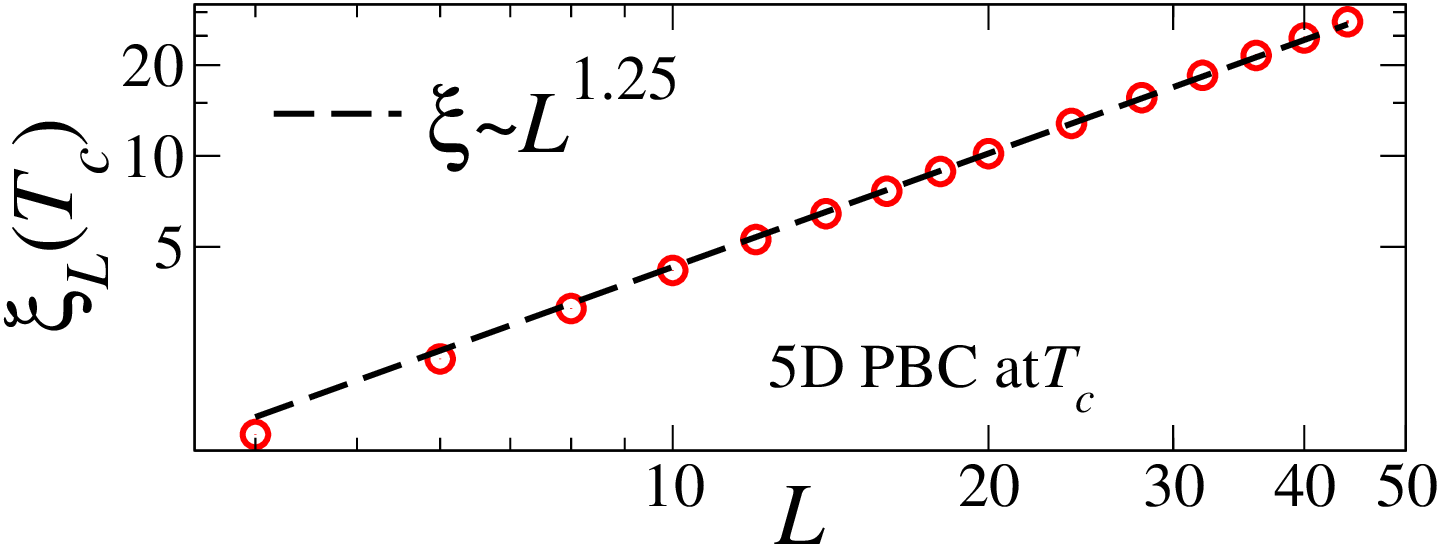}\vskip 5mm
  
  \includegraphics[width=0.8\columnwidth]{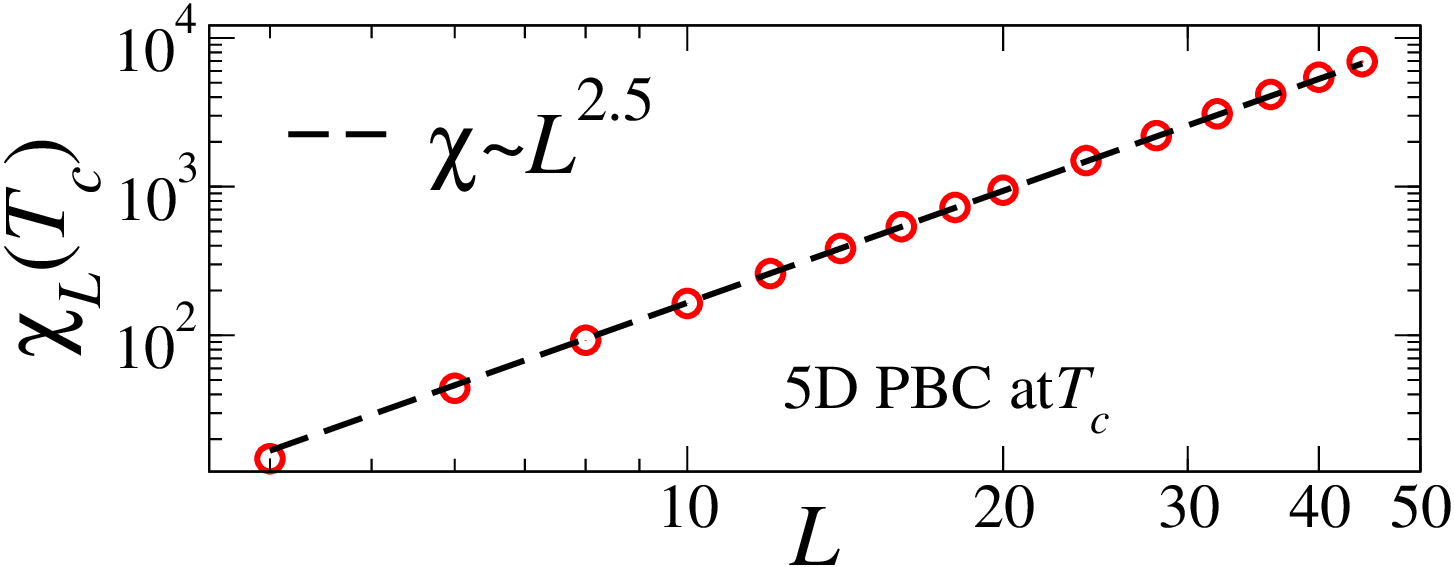}\vskip 5mm
  
  \includegraphics[width=0.8\columnwidth]{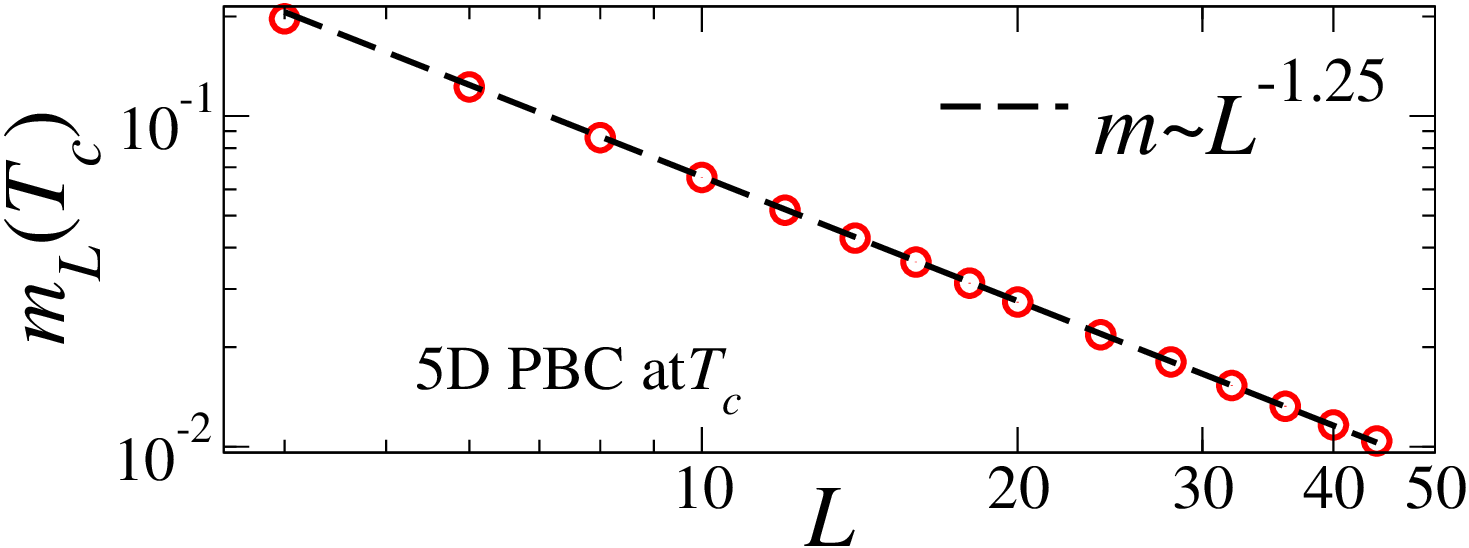}
\caption{(a) Correlation length, (b) susceptibility and (c) magnetization for the 5D PBC Ising model at the extrapolated 
critical point $T_c$.  These  agree with Eqs.(\ref{q}), (\ref{FSSchi}) and (\ref{FSSm}) 
(dashed lines) with $q=d/d_c = 5/4$.}
  \label{Chi_and_Xi_vs_L_5dPBC_Tc}
  \end{center}
\end{figure}

\begin{figure}
  \begin{center}
  \includegraphics[width=0.8\columnwidth]{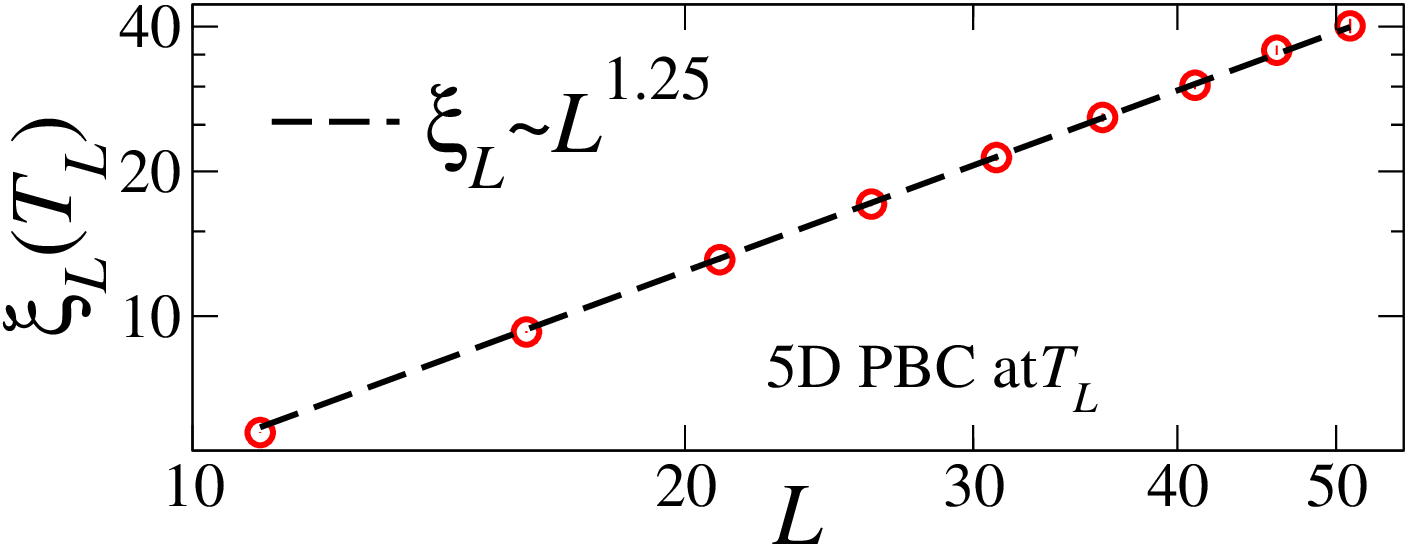}\vskip 5mm
  
  \includegraphics[width=0.8\columnwidth]{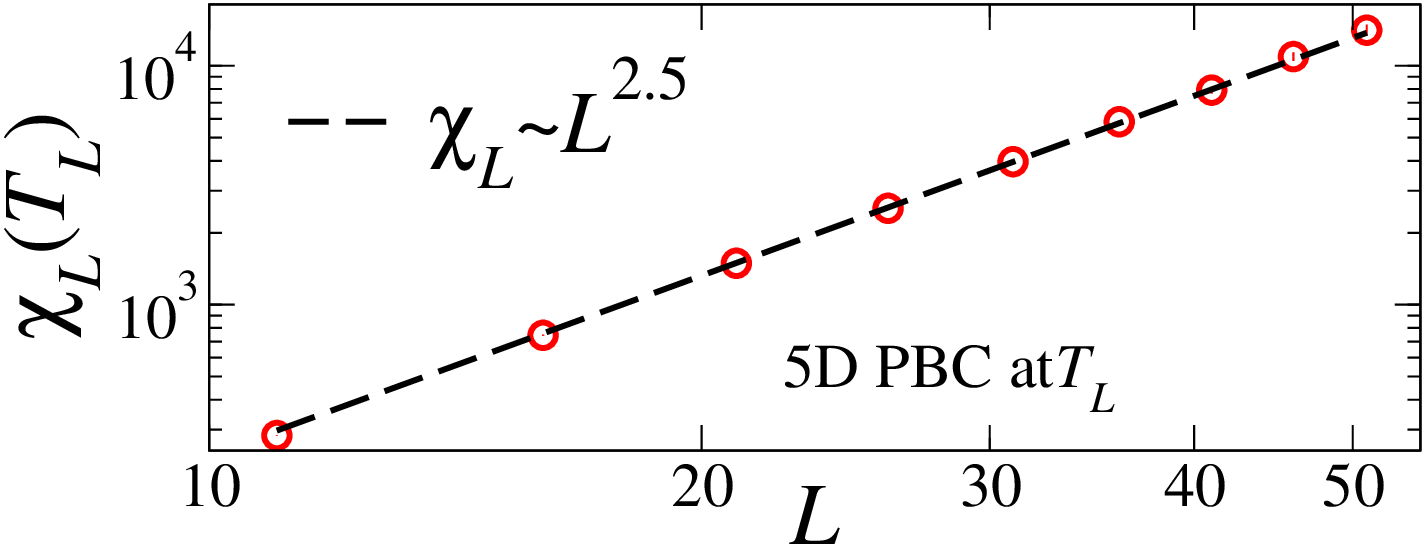}\vskip 5mm
  
  \includegraphics[width=0.8\columnwidth]{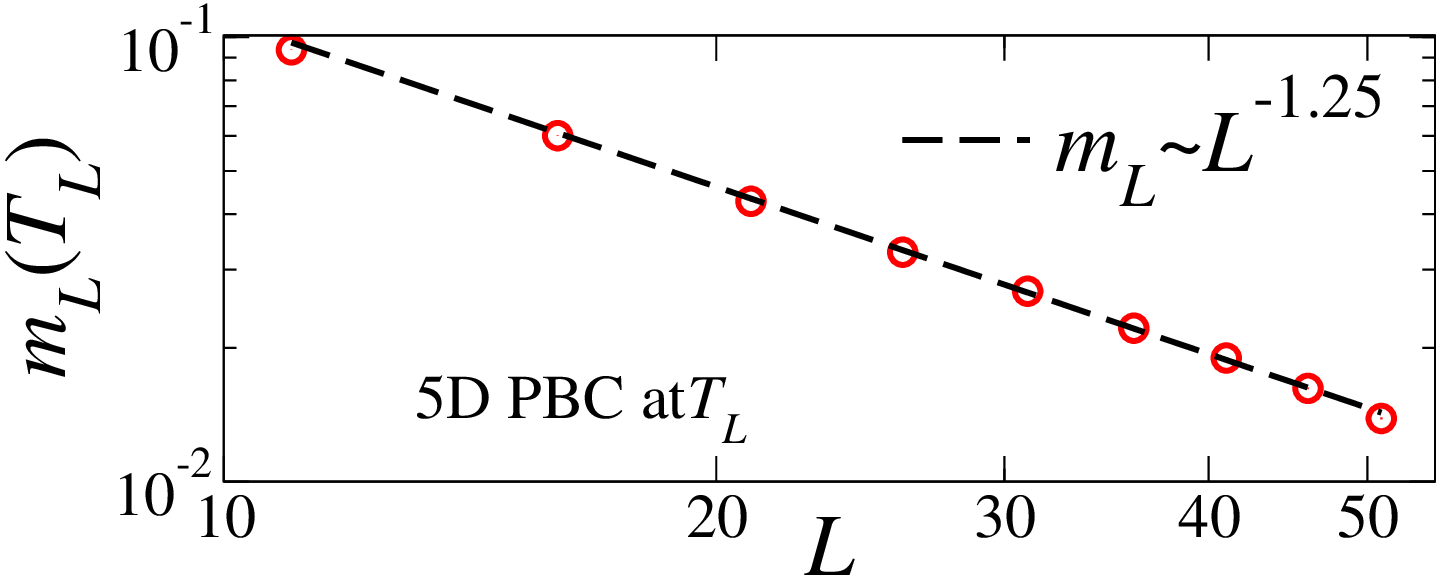}
\caption{(a) Correlation length, (b) susceptibility and (c) magnetization for the 5D PBC Ising model at the pseudocritical point $T_L$. These also agree with 
Eqs.(\ref{q}), (\ref{FSSchi}) and (\ref{FSSm})  (dashed lines), with $q=d/d_c = 5/4$.}
  \label{Chi_and_Xi_vs_L_5dPBC_TcL}
  \end{center}
\end{figure}

\begin{figure}
  \begin{center}
  \includegraphics[width=0.8\columnwidth]{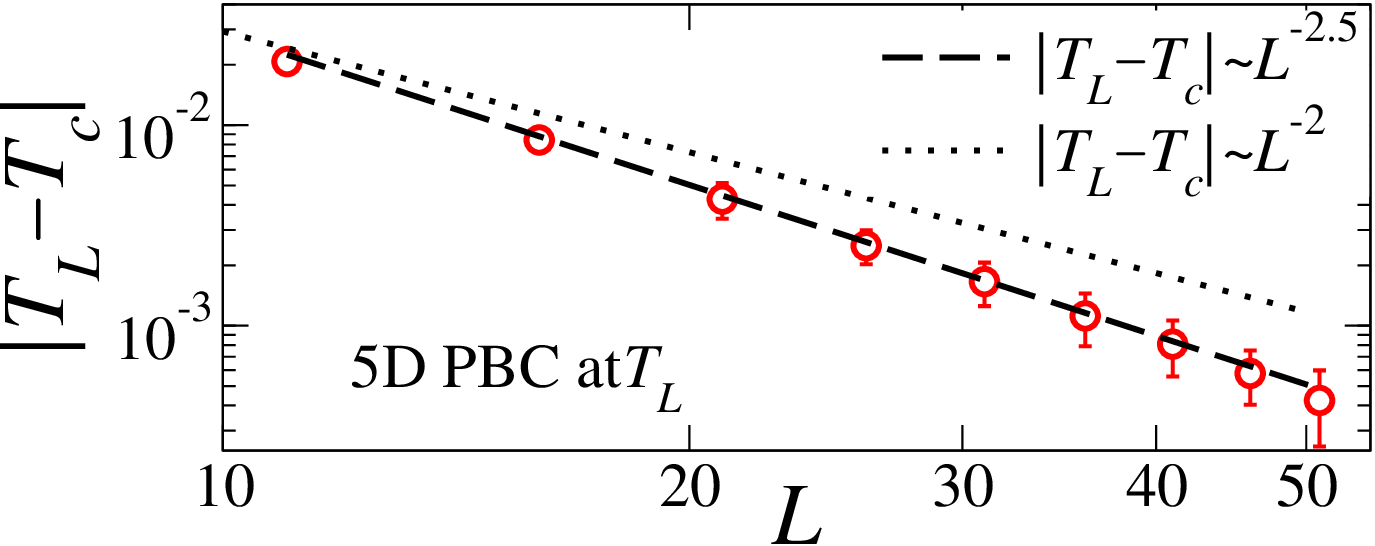}\vskip 5mm
  
  \includegraphics[width=0.8\columnwidth]{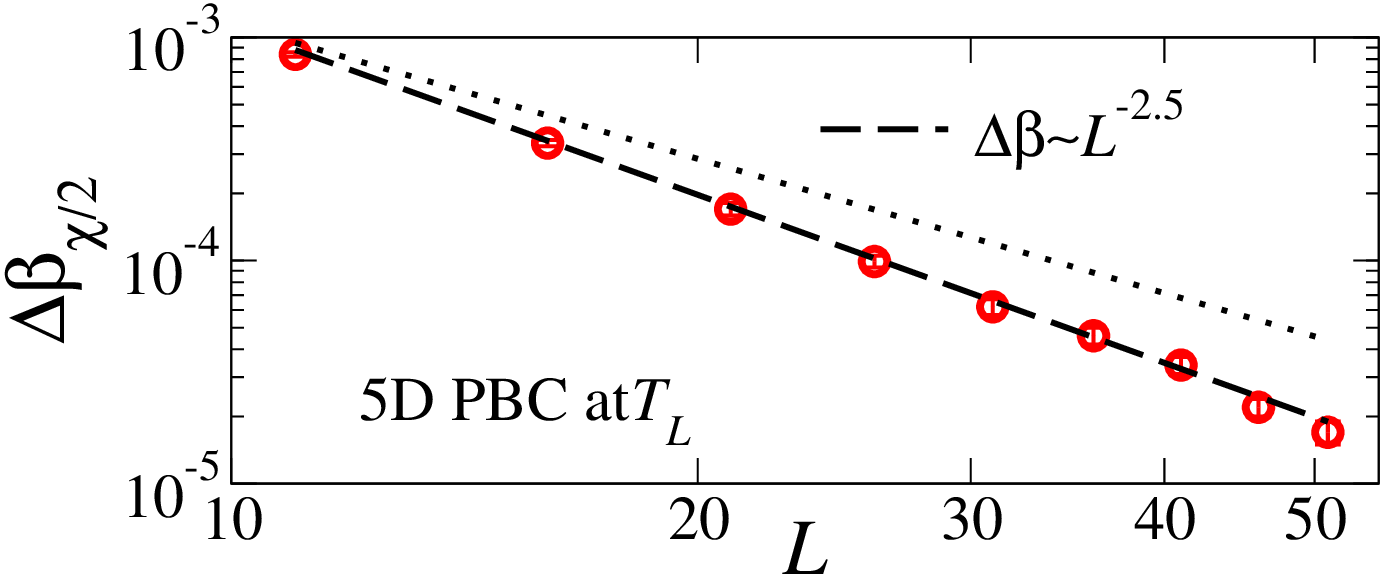}\vskip 5mm
  
  \includegraphics[width=0.8\columnwidth]{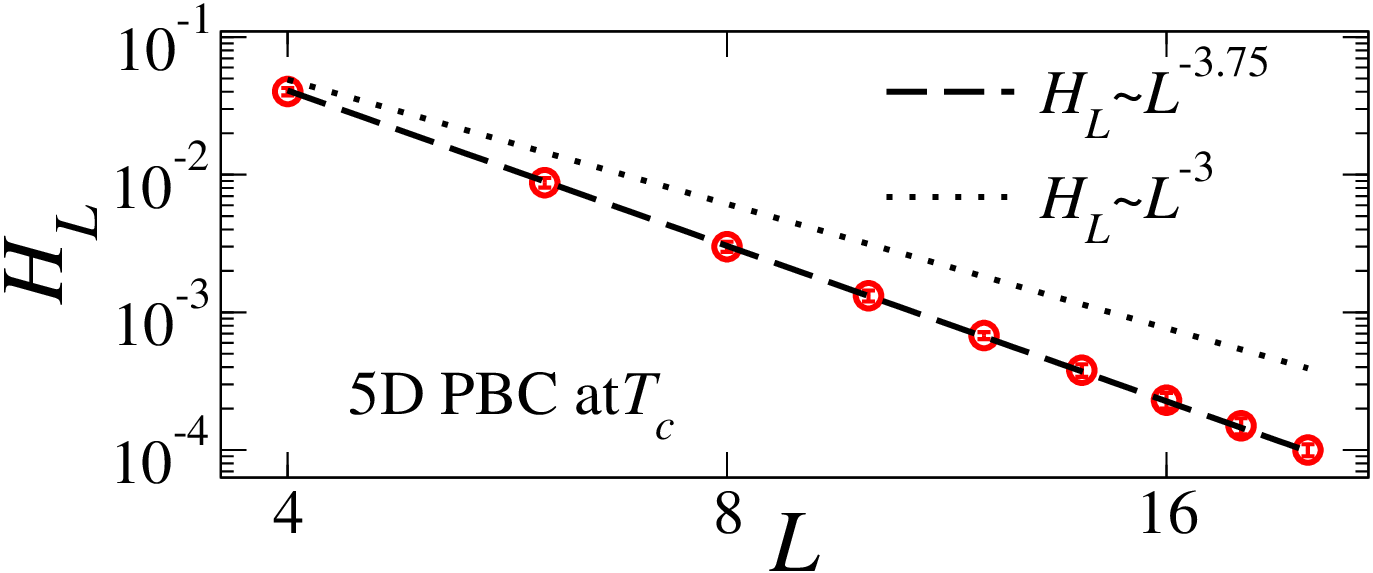}
  \caption{
(a) Scaling of the shift in temperature from the critical point $T_c$ to the pseudocritical point $T_L$  for the susceptibility with PBC's. 
The shift exponent is in good agreement with Eq.(\ref{shiftround3}) with $q=5/4$. 
(b) The rounding exponent defined through the width at half-height of the susceptibility is also supportive of  Eq.(\ref{shiftround3}) with $q=5/4$ for  PBC's.
(Here $\Delta \beta$ is the width in units of $\beta = 1/T$.)
(c) The shift of the pseudocritical field favours  $H_L \propto L^{-\lambda_H}$ with the $Q$-FSS value 
 $\lambda_H = q\beta \delta / \nu = 3.75$ instead
 of the Gaussian value $\lambda_H = \beta \delta / \nu = 3$. This gives evidence for $q$ associated with the odd sector of the model too. 
 Dashed lines  correspond to expected power laws with $Q$-FSS while dotted lines correspond to Gaussian FSS.
   }
  \label{Shift_T_and_h_vs_L_forChi_5dPBC}
  \end{center}
\end{figure}

We begin our report with an analysis of FSS at the critical point and the pseudocritical point for PBC's.
In Fig.\ref{Chi_and_Xi_vs_L_5dPBC_Tc}, the FSS for the correlation length, susceptibility and magnetization are plotted on logarithmic scales. 
The same quantities are plotted at the pseudocritical point in  Fig.\ref{Chi_and_Xi_vs_L_5dPBC_TcL}. 
The modified, or $Q$-FSS formulae (\ref{q}), (\ref{FSSchi}) and (\ref{FSSm}) are re-verified with $q=d/d_c = 5/4$ instead of the Gaussian $q=1$ in each case.
(Here, and elsewhere, rather than presenting fits to the numerical data, we show the expected slopes of the log-log plots, our aim being to discriminate between two scenarios for which critical exponents are already known, rather than to determine critical exponents numerically.)
This is in agreement with 
Refs.\cite{Br82,Bi85,BNPY,BZJ85,Binder87,RiNi94,Mon1996,PaRu96,LuBl97a,LuBl97b,BlLu97c,Luijtenthesis,LuBi99,BiLuquote,Turks,JoYo05,LuMa11,BDT,Review} for PBC's. 
In Fig.\ref{Shift_T_and_h_vs_L_forChi_5dPBC}(a) and (b), we also test the FSS of the pseudocritical temperature and associated rounding.
Each of these deliver results consistent with $q=d/d_c$ in Eq.(\ref{shiftround3}). 
Fig.\ref{Shift_T_and_h_vs_L_forChi_5dPBC}(c) depicts the FSS of the pseudocritical field. The result is in excellent agreement with $q=d/d_c$ and rules out the Gaussian $q=1$, indicating that $q$ is associated with the odd sector of the model (the second argument on the right-hand side of Eq.(\ref{BNPY}))
as well as the even one (the first argument in Eq.(\ref{BNPY})).

These results reconfirm that $q=d/d_c = 5/4$ governs modified FSS above the upper critical dimension with PBC's, results which can be traced back to dangerous irrelevant variables in the RG picture. 
The standard expectation is that $q=1$ in the FBC case even above $d_c=4$ dimensions \cite{JoYo05}.
We next move on to test that scenario. We will see that the standard picture is incorrect.

\subsection{Free boundary conditions}

It is reported in Ref.\cite{RuGa85} that Eq.(\ref{FSSchi}) ``cannot hold for free boundary conditions because it lies above a strict upper bound'' (namely $L^{\gamma/\nu}=L^2$) established in Ref.\cite{Watson} (see also Ref.\cite{Gunton}). 
However the Fourier analysis of Ref.\cite{RuGa85}, which yielded the same conclusions, neglects the quartic part of the $\phi^4$ action because of an expectation that the Gaussian result should apply to leading order. 
Moreover,  the  bound was determined   at criticality instead of  at pseudocriticality.
It was shown in Refs.\cite{LuBl97a,LuBl97b,BlLu97c} that for PBC's the  FSS behaviour (\ref{FSSchi}) is obtained from  precisely this interaction term. 

In Ref.~\cite{JoYo05}, it is stated that for FBC's ``it seems obvious that even for $d > 4$ the behavior of the system will be affected
when $\xi_L$ becomes of order $L$'', rather than the larger $L^{d/4}$. 
On this basis, it was stated that the standard FSS expressions (corresponding to $q=1$) are expected to apply. 
The larger length scale $L^{d/4}$ was only expected to contribute to corrections to scaling in some unspecified manner.
The numerical results of Ref.~\cite{LuMa11} would initially appear to substantiate these conclusions and speculations.
There, the Ising model was simulated on 5D FBC lattices with up to $L=24$ sites in each direction.

However, for free boundary conditions, the number of sites in the bulk is $(L-2)^d$ and the number on a $(d-m)$-dimensional sub-manifold is $2^m {d \choose m} 
(L-2)^{d-m}$.
Therefore the proportion of sites in the bulk of a size-$L$ lattice is  $(1 - 2/L)^d$, the remaining ones being in lower-dimensional manifolds. 
Thus small FBC lattices, although defined in five dimensions, may not genuinely represent a five-dimensional Ising system, and to achieve this, either enormous lattices must be used or another trick has to be introduced.
In the former case, to have 1\% of sites outside the bulk with free boundary conditions, for example, would require a lattice of linear extent $L\approx 10^3$ in $d=5$ dimensions. 
In this case,  the number of lattice points to simulate would be $10^{15}$. 
At the moment, simulating $L\approx 50$ ($\sim 2 \times 10^8$ sites) is about the limit which we can achieve and such lattices have only about 60-70\% of points in the bulk.
Therefore free boundary conditions manifest significant corrections coming from lattice sites on lower-dimensional manifolds, corrections which will reduce as larger $L$-values become accessible. 
The $L=4$ to $L=24$, 5D lattices of Ref.~\cite{LuMa11} have only between $3\%$ and $65\%$ of sites in the bulk and do not represent five-dimensionality. 
The resulting  conclusion that $\chi_L \sim L^2$ is therefore not a 5D one.

To ameliorate the problem with available computing power and to probe the dimensionality of the system, we remove the contributions of sites close to the surfaces.
Specifically, we simulated FBC Ising lattices  up to $L = 40$, removing the contribution of $L/4$ sites at each boundary,  resulting in lattice cores of size $L_{\rm core}=L/2$.  
For $L=16$, for instance, the contributions from the four sites closest to each boundary are removed and the remaining $8^d$ sites used for the average (\ref{chinumeric}).
A similar tactic of removing the contributions of boundary spins has been used to examine the Ising model on a negative-curvature substrate \cite{pseudosphere}.

\begin{figure}
  \begin{center}
  \includegraphics[width=0.8\columnwidth]{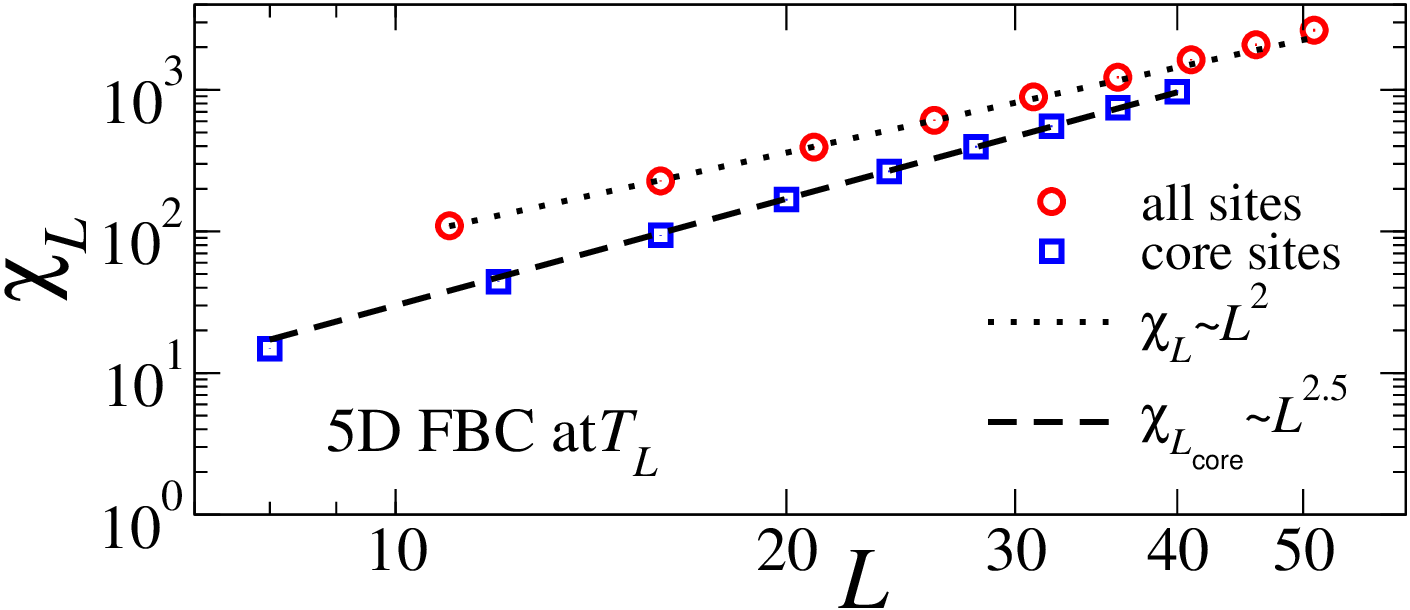}\vskip 5mm
  
  \includegraphics[width=0.82\columnwidth]{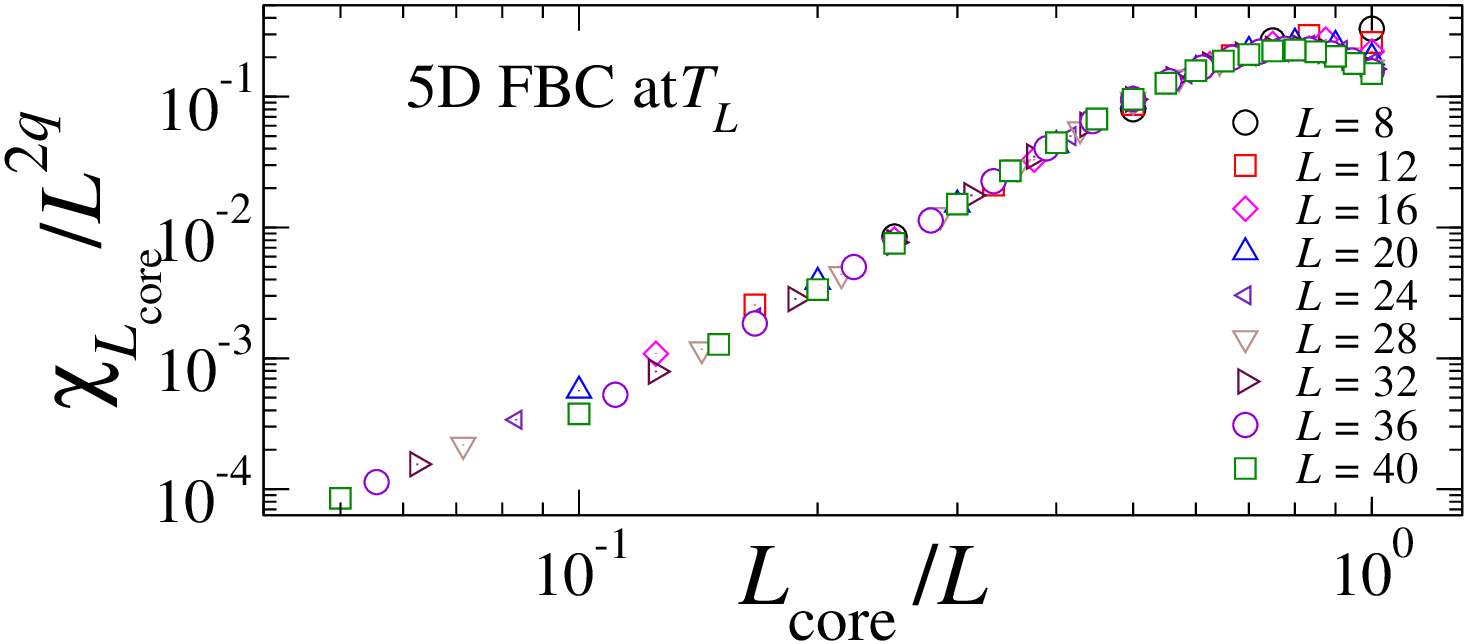}\vskip 5mm
  
  \includegraphics[width=0.8\columnwidth]{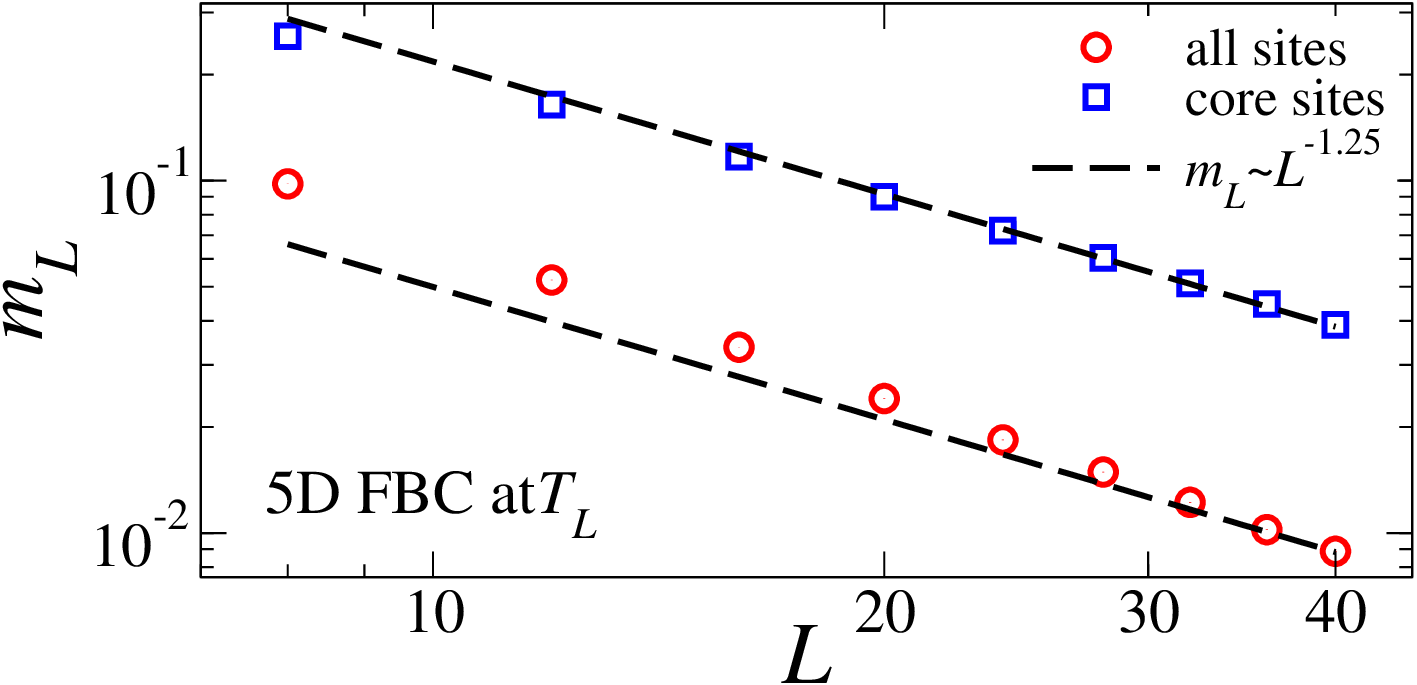}
  \caption{FSS at the pseudocritical point: 
  In panel (a) the top curve represents $\chi_L(T_L)$ with all the sites. The apparent Gaussian FSS $\chi_L \sim L^2$ (dotted line) 
is due to the 4D lattice surfaces. For larger $L$ values, the data points start to deviate from the curve. 
The bottom curve shows $\chi_L \sim L^{5/2}$ (dashed line)
 when peripheral sites are removed.
 (b)  $Q$-FSS is followed for a broad range of $L_{\rm core}/L$, and not only for $1/2$. 
 The different curves (coloured online) correspond to different underlying $L$-values.
 (c) Magnetization from the core sites versus $L$ for FBC's at pseudocriticality also displays $Q$-FSS (dashed line). 
  }
  \label{Chi_vs_L_5dFBC}
  \end{center}
\end{figure}

We start the FBC analysis with Fig.\ref{Chi_vs_L_5dFBC}(a) which depicts the pseudocritical susceptibility plotted against $L$.
In the upper curve, all lattice sites -- including those on the lattice boundary -- contribute to the calculation of the susceptibility.
For small lattice sizes, there appears to be good agreement with the Gaussian $\chi_L \sim L^2$. 
However, as the lattice sizes increase, the plot deviates from this apparent match. 
We interpret the match as being spurious - it is due to the fact that the full susceptibility  is not, in fact, genuinely representative of a five-dimensional system.
This is because most of the sites belong to a four- (or fewer) dimensional submanifold.
The deviation from Gaussian behaviour is due to the onset of genuine five-dimensionality as the lattices grow.

This conclusion is reinforced in the lower curve. There the contributions to the susceptibility of half of the sites close to the boundaries have been removed. 
The remaining lattice core is genuinely five-dimensional. 
The fit is clearly consistent with the $Q$-FSS prediction $\chi_L \sim L^{2q} = L^{5/2}$. 
This contradicts  the standard wisdom discussed above.

In Fig.\ref{Chi_vs_L_5dFBC}(a), the lattice core is of length $L_c = L/2$. 
In Fig.\ref{Chi_vs_L_5dFBC}(b), different values of the ratio $L_c/L$ are used. 
That plot indicates that the core susceptibility is linearly related to $L^{2q}$ for a range of values of the core-to-size ratio, so that Fig.\ref{Chi_vs_L_5dFBC}(a) is not an artefact of the choice $L_c/L=1/2$.

Fig.\ref{Chi_vs_L_5dFBC}(c) contains analogous plots for the magnetization. 
The upper curve displays the pseudocritical core magnetization and follows the modified or $Q$-FSS behaviour $M\sim L^{-q}$. 
The lower curve contains the full lattice magnetization and is neither Gaussian nor $Q$. 
These results  signal that  the exponent $q$ governing $Q$-FSS  may in fact be universal  at pseudocriticality.

\begin{figure}
  \begin{center}
  \includegraphics[width=0.8\columnwidth]{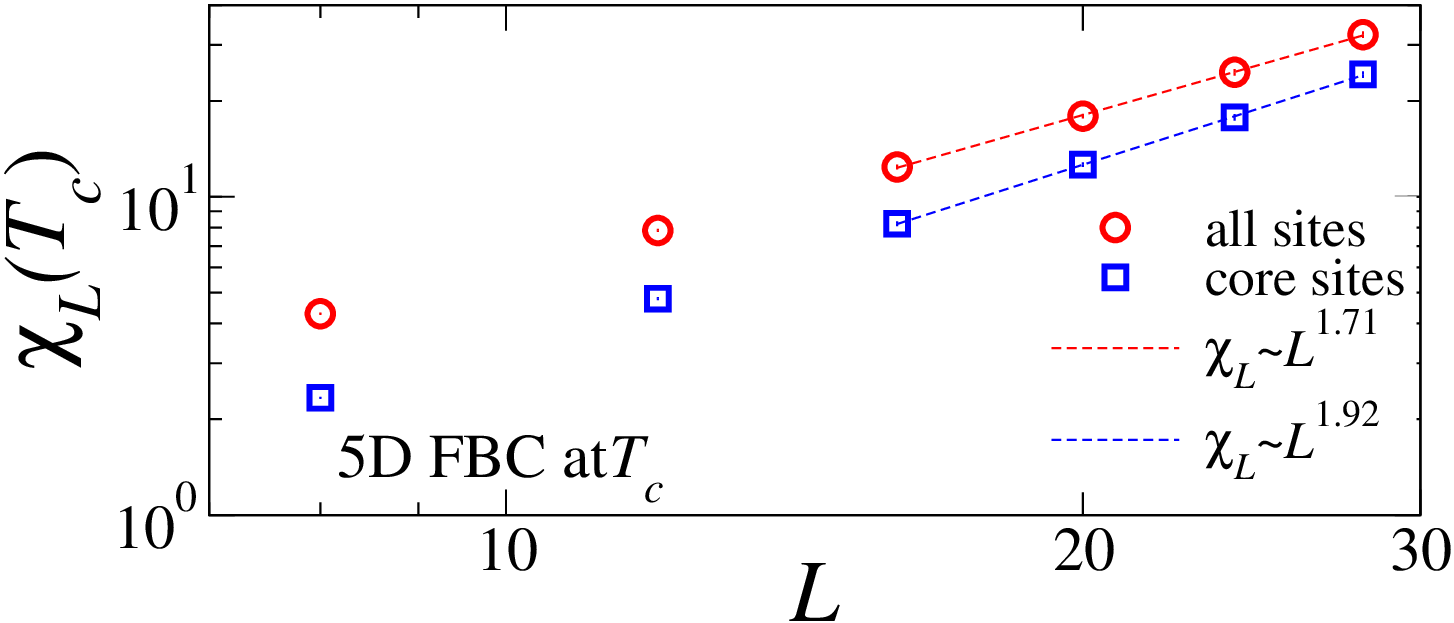}
  
  \includegraphics[width=0.8\columnwidth]{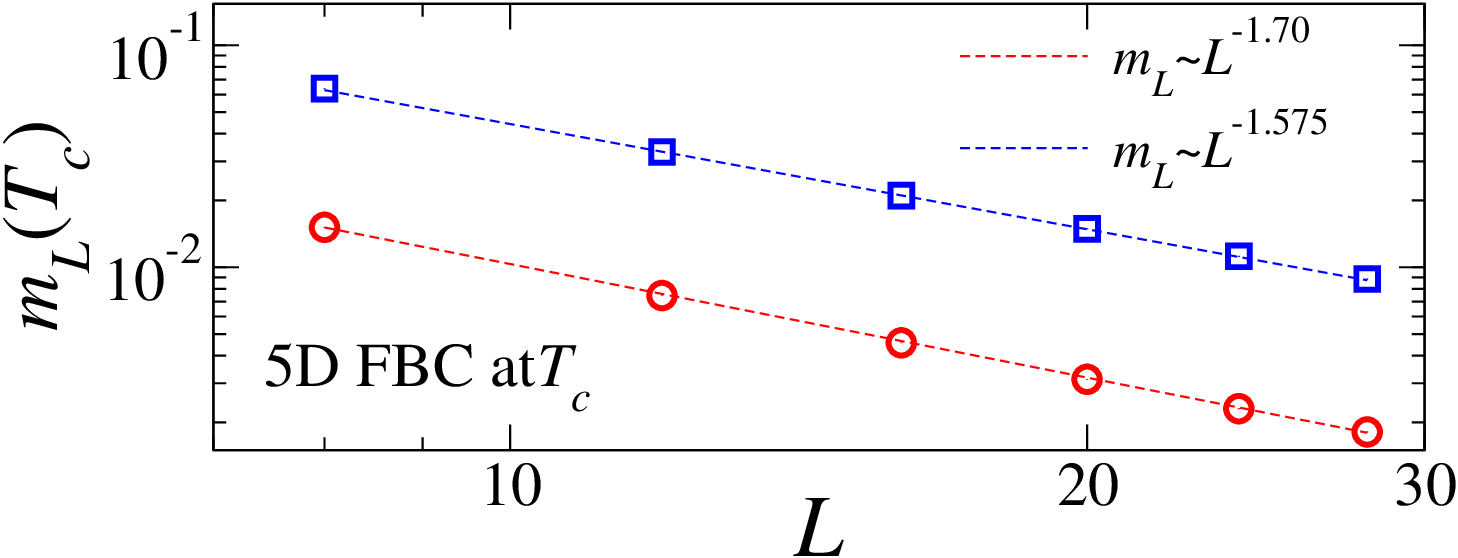}
  \caption{
(a) FSS at $T_c$ for 5D FBC lattices. The upper curve represents the  calculation using all the sites and the fitted curve (red online) is  $\chi_L \sim L^{1.71(2)}$.
The calculation corresponding to the lower curve has the outer half of the sites removed and a fit gives an exponent $\chi_L \sim L^{1.92(2)}$ (blue online).
(b)
A fit to the magnetization for FBC's at $T_L$ gives  $m_L \sim L^{-1.700(15)}$ with all sites 
and $m_L \sim L^{-1.575(15)}$ using only the core sites. 
Neither corresponds to the Gaussian  $m_L \sim L^{-1}$ or the $Q$-form $m_L \sim L^{-1.25}$.
}
  \label{Chi_vs_L_5dFBC_TcM_vs_L_5dFBC}
  \end{center}
\end{figure}

We next analyse FSS at the infinite-volume critical point with FBC's. 
In Fig.\ref{Chi_vs_L_5dFBC_TcM_vs_L_5dFBC}(a), a fit to the critical susceptibility calculated using all sites of each FBC lattice (upper curve) 
delivers an exponent $1.71(2)$. This is different to the Gaussian value $\gamma/\nu =2$ and the $Q$-value $q \gamma /\nu  = 2.5$ which we witnessed at the pseudocritical point.
The estimate for this quantity from Ref.~\cite{LuMa11} was compatible with $\gamma/\nu =2$ but using far smaller lattices (and using the estimate $0.113\,913\,9$ for $\beta_c$ instead of our $0.113\,915\,5$).
The lower curve in the panel shows the result of the calculation with the outer half of the sites removed. 
A fit gives an exponent $1.92(2)$, still far from the $Q$-value and incompatible with the Gaussian value too.

Fig.\ref{Chi_vs_L_5dFBC_TcM_vs_L_5dFBC}(b)  contains analogous fits for the magnetization using all lattice sites (lower curve) and with peripheral sites removed (upper curve). 
Again, at $T_c$, the data follow neither the Gaussian nor the $Q$-behaviour. 
Indeed, a fit using all points gives an exponent $-1.70(2)$ with all sites 
and $-1.58(2)$ using only the core sites and these are both far from the Gaussian expectation  $-\beta / \nu = -1$ and the $Q$ expectation $- q \beta /\nu = -1.25$.

We interpret these results as indicating that FBC lattices at criticality with all sites or even only core sites  are too far from the pseudocritical point to be inside the FSS regime. 
To further explore this interpretation, we next examine rounding and shifting.

If the shifting is bigger than the rounding,
the critical point $T_c$ will be too far from the pseudocritical point $T_L$ to "feel" the FSS peak -- it will
be outside the FSS zone. This would explain why, even 
if quantities at the pseudocritical point scale as $Q$-FSS,
the same quantities at the critical point may scale differently, at least for the lattice sizes we can deal with. 
This would reconcile $\chi_L \sim L^{d/2}$ at $T_L$ with some other behaviour at $T_c$. 
(However, we remind that the discussion above indicates that this other behaviour is not Gaussian.)

The corresponding rounding and shifting plots are contained in Fig.\ref{rounding_T_vs_L_forChi_5dPBCShift_T_vs_L_forChi_5dFBCCumBinFBCPBC}.
Defining the rounding exponent from the width at half-height of the susceptibility, Eq.(\ref{shiftround3}) is verified for the 
rounding exponent. 
However, instead of the shift exponent being derived from $Q$-FSS in the FBC case (corresponding to Eq.(\ref{shiftround3})),
we find  {Gaussian-like behaviour (corresponding to Eq.(\ref{shiftround2})).
Removing the peripheral sites doesn't change the location of the pseudocritical point within error bars, so the shift exponent for the core lattice is essentially the same as for the entire lattice
 -- see Fig.\ref{Shift_T_vs_L_forChi_5dFBC_2}.

We conclude that, when the peripheral sites are removed, all pseudocritical behaviour in the FBC $d>d_c$ case is driven by $Q$-FSS except that of the shift exponent, which is more akin to Gaussian behaviour.
This anomalous behaviour of the shift exponent should not be entirely unexpected as similar anomalies are observed even in two-dimensional lattices 
 with various boundary conditions \cite{JaKe02}.

\begin{figure}
  \begin{center}
\includegraphics[width=0.8\columnwidth]{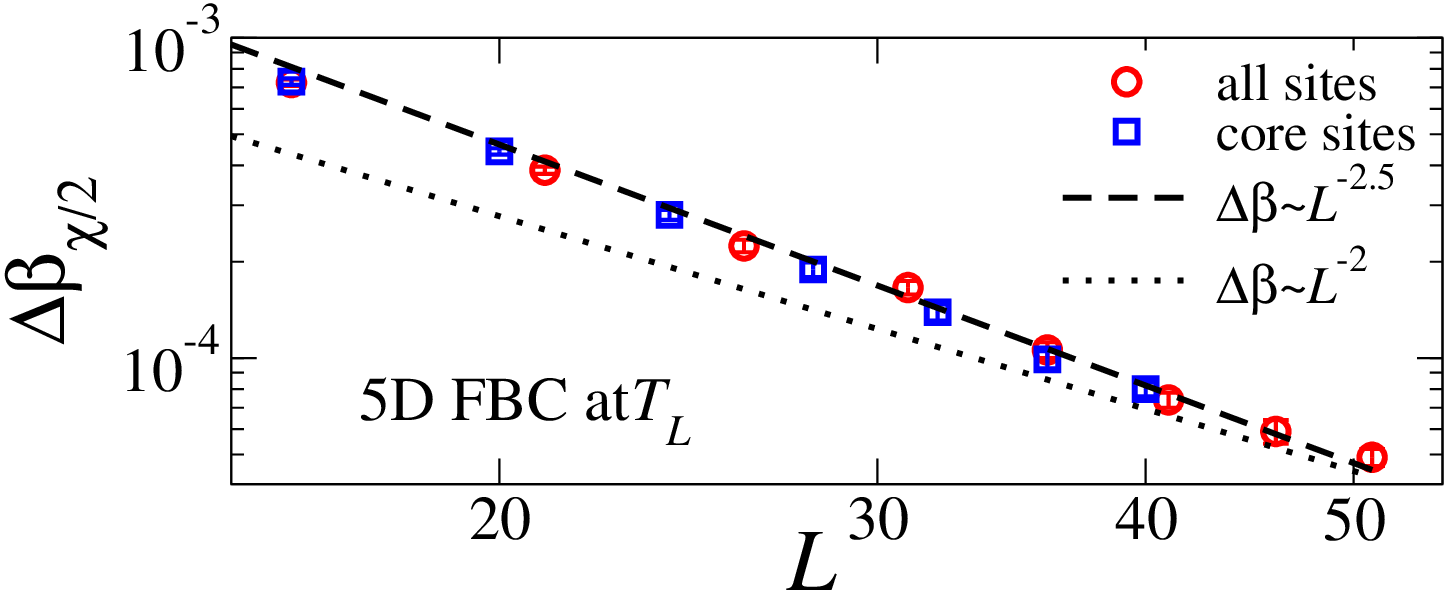}

\includegraphics[width=0.8\columnwidth]{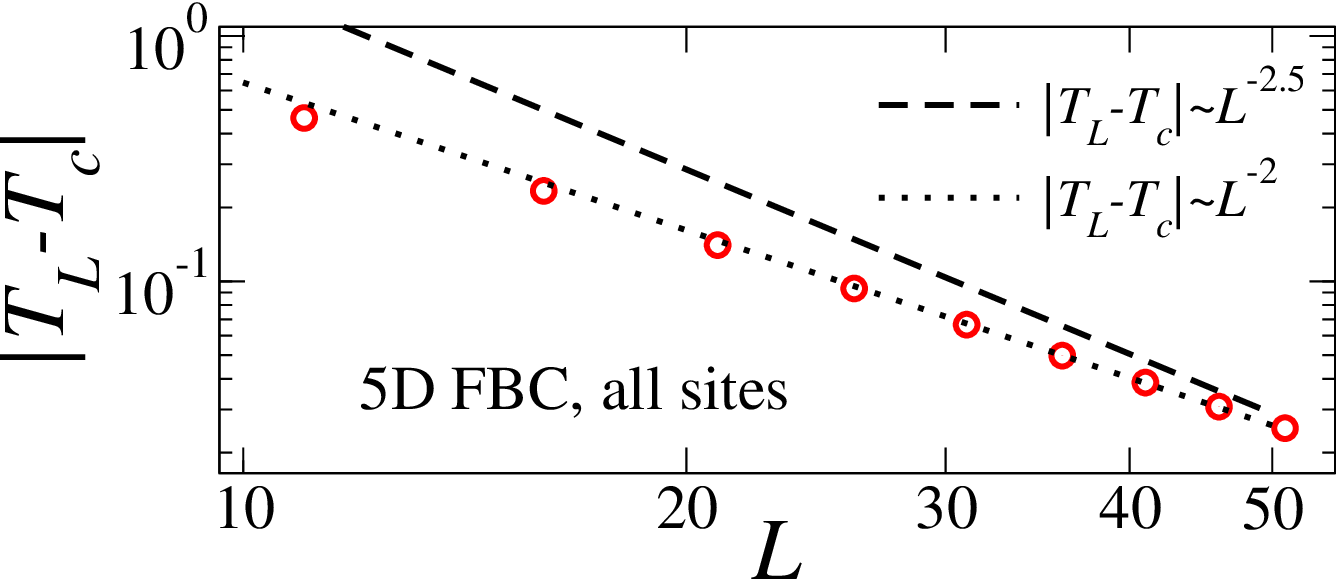}  

\includegraphics[width=0.8\columnwidth]{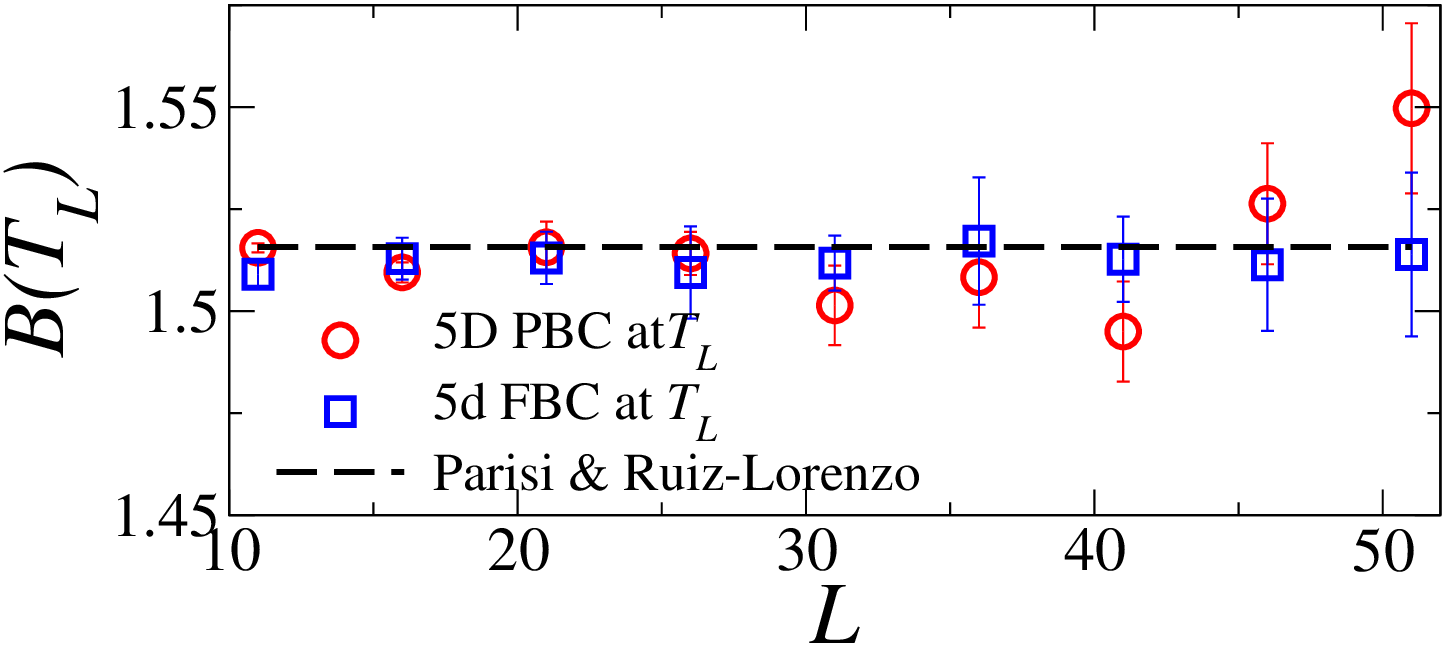}
  \caption{
  (a) The rounding exponent defined through the width at half-height of the susceptibility for FBC's 
is in agreement with the $Q$-value of Eq.(\ref{shiftround2}) (dashed line). 
(b) The shifting exponent for FBC's from the susceptibility peak
using all sites corresponds to the 
Gaussian equation (\ref{shiftround3}) (dotted line). 
 Removing peripheral sites doesn't change the shift $T_L-T_c$ within errors (see Fig.\ref{Shift_T_vs_L_forChi_5dFBC_2}).
(c) The Binder cumulant for PBC's and FBC's at the pseudocritical point are both in good agreement with the value of Parisi 
and Ruiz-Lorenzo \cite{PaRu96}.}
  \label{rounding_T_vs_L_forChi_5dPBCShift_T_vs_L_forChi_5dFBCCumBinFBCPBC}
  \end{center}
\end{figure}

\begin{figure}
  \begin{center}
  \includegraphics[width=0.8\columnwidth]{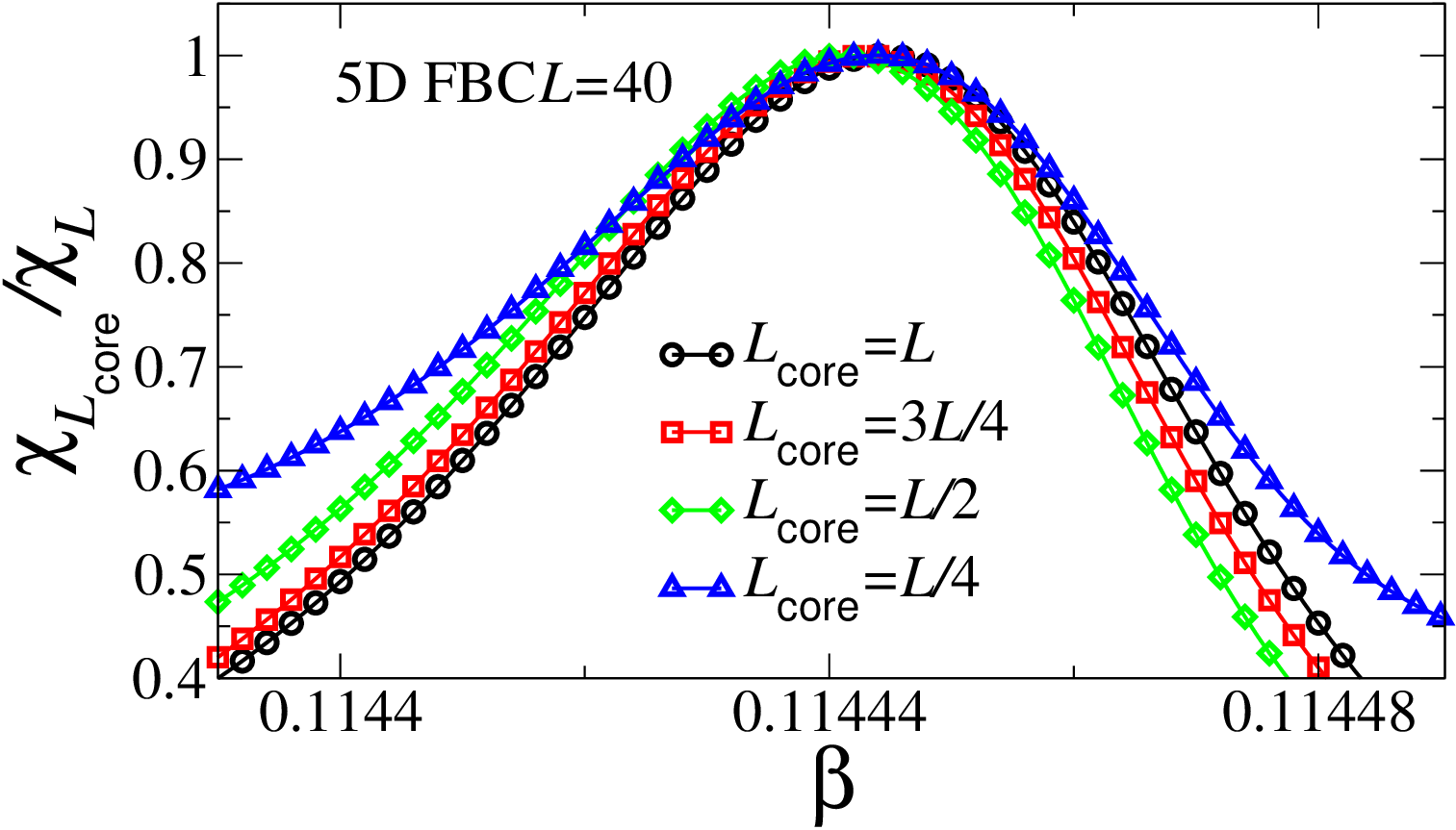}
  \caption{
Removing sites doesn't change the shift in temperature for the 5D Ising model with  FBC's.
}
  \label{Shift_T_vs_L_forChi_5dFBC_2}
  \end{center}
\end{figure}


Finally, more support for universality of FSS at the pseudocritical point comes from  Fig.\ref{rounding_T_vs_L_forChi_5dPBCShift_T_vs_L_forChi_5dFBCCumBinFBCPBC}(c), in which we plot the Binder cumulant \cite{Schladming} at the pseudocritical point for both the PBC and FBC cases.
The result, including for FBC's, is in good agreement with that coming from Parisi and Ruiz-Lorenzo's analysis, which was for PBC's \cite{PaRu96}.

\section{Analytic interpretation}

To summarize so far, systems with PBC's above the upper critical dimension are long known to obey modified (what we call ``$Q$''-) FSS
involving the exponent $q=d/d_c$ both at the infinite-volume critical point and at the finite-volume pseudocritical point. 
Until now, it has been believed that the equivalent systems with FBC's have Gaussian FSS (corresponding to $q=1$), although there is no analytical proof for this and the limited numerics that exist are for lattices whose lower-dimensional boundaries exert very strong influence.
We have shown that once the lower-dimensional influence of the peripheries is removed, FBC systems in fact also exhibit $Q$-FSS at pseudocriticality. Using the same technique at the infinite volume critical point, we find no evidence for either Gaussian FSS or for $Q$-FSS.
Because the rounding is smaller than the shifting, we attribute this to the fact that $T_c$ is too far from $T_L$ to come under the influence of FSS there. 
This means that (a) the conventional  FSS paradigm for FBC lattices above $d_c$  is {\emph{unsupported}}, and this is particularly clear at pseudocriticality \cite{RuGa85,LuMa11,Watson,Gunton} and (b) it offers evidence for the likely universality of $q$ at pseudocriticality.

That $q$ is indeed $d/d_c$ and not unity can be seen from the Lee-Yang zeros of the partition function \cite{LY}. 
For finite systems with discrete energy levels as we have here, the partition function becomes a polynomial in $\exp{(h)}$.
Its zeros form a discrete set, distributed on the imaginary-$h$ axis \cite{LY} and labelled $h_j(L,t)$.
Here $j$ is an integer index whose value is $1$ for the zero closest to the real-$h$ axis and increases as one moves out along their locus.
The Lee-Yang zeros condense in the thermodynamic limit, where the counterpart of $h_1$ is the Lee-Yang edge, $h_{\rm{edge}}(t)$.
The standard scaling behaviour for the edge of the distribution is ${{ h_{\rm{edge}}(t) \sim t^\Delta }}$ where $\Delta = \beta \delta$
is the gap exponent.
The $Q$-FSS for the $j$th zero is  
\begin{equation}
 h_j(L) \sim  \left({\frac{j}{L^d}}\right)^{\frac{\Delta q}{\nu d}}
 \label{hjL}
\end{equation}
 where $\Delta = 3/2$ above $d=d_c=4$ 
 \cite{IPZ}.
As usual, Gaussian FSS is recovered simply by setting $q=1$.
Following Ref.\cite{KJJ2006},  the finite-size susceptibility may be written in terms of the Lee-Yang zeros as
\begin{equation}
 \chi_L \sim  L^{-d} \sum_{j=1}^{L^d}{{h_j^{-2}(L)}}
,
\end{equation}
which,  gives
\begin{equation}
 \chi_L \sim L^{\frac{2\Delta q}{\nu}-d}
 \sum_{j=1}^{L^d}{j^{-\frac{2\Delta q}{\nu d}}}
  .
\label{21}
\end{equation}
Together with  the  static scaling relations 
$ 2\beta + \gamma  =  2 - \alpha$ and 
$ \beta (\delta - 1)   = \gamma$,  
matching Eq.(\ref{FSSchi}) to Eq.(\ref{21})  gives 
\begin{equation}
 \frac{\nu d}{q} = 2 - \alpha .
\label{hyperhyperscaling}
\end{equation}
This recovers  standard hyperscaling  (\ref{hyperscaling})   provided $q=1$ below $d_c$.
However, if $q=1$ persisted above $d_c$, the summation on the right-hand side of  Eq.(\ref{21}) would deliver a spurious leading logarithm in the $d=6$ Ising case, {\emph{irrespective of boundary conditions}}. 
It is well known that such leading logarithms can occur {\emph{only}} at the upper critical dimension \cite{Br82,KJJ2006,WeRi73,ourhighD}. Indeed, Butera and Pernici have recently given convincing evidence of this for Ising and scalar-field models using high-temperature series expansions in the thermodynamic limit -- i.e., there are no leading multiplicative logarithmic corrections in 6D \cite{Butera}.
This indicates that $q \ne 1$ {\emph{even for FBC's}}. The incorporation of $q$ into Eq.(\ref{hyperhyperscaling})
 extends hyperscaling  to $d>d_c$.

The validity of this picture for both PBC's and FBC's is confirmed in Fig.\ref{LY0_vs_L_forChi_5dPBC}.
There the FSS for the first two Lee-Yang zeros for PBC lattices at criticality and pseudocriticality confirm Eq.(\ref{hjL}) with $q=5/4$, i.e., 
the edge scales as $h_1 \sim L^{-15/4}$ instead of $h_1 \sim L^{-3}$.
The same behaviour (with $q=5/4$) is confirmed in the FBC case at pseudocriticality using all lattice sites as well as using lattice cores.

\begin{figure}
  \begin{center}
  \includegraphics[width=0.8\columnwidth]{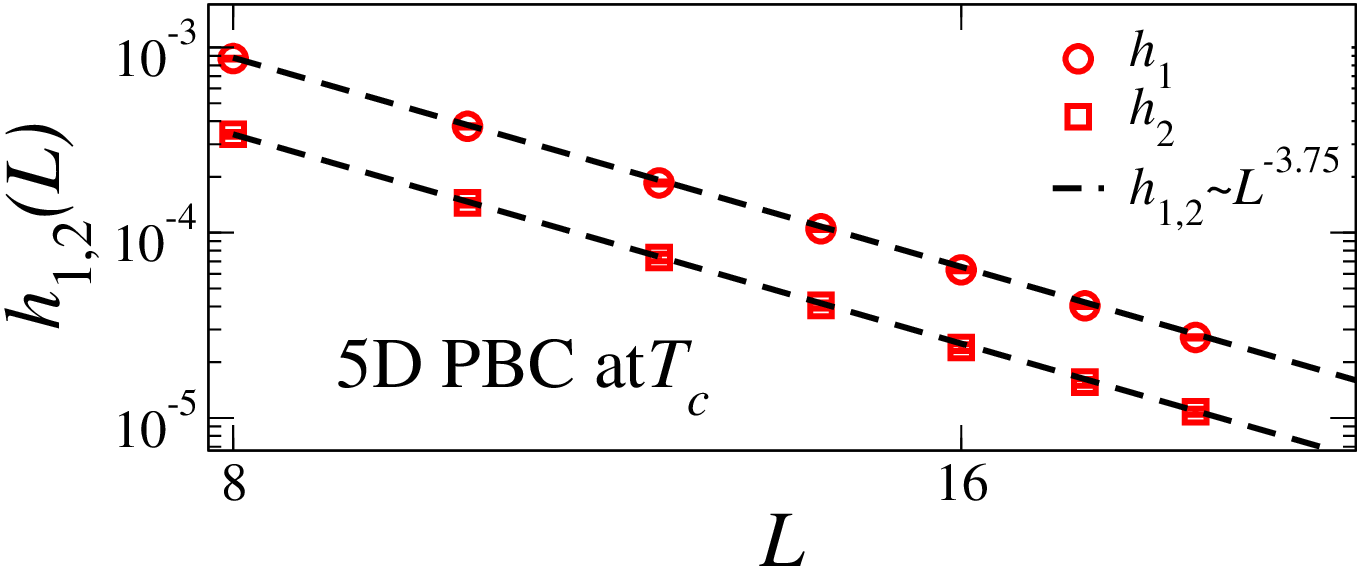}
  \includegraphics[width=0.8\columnwidth]{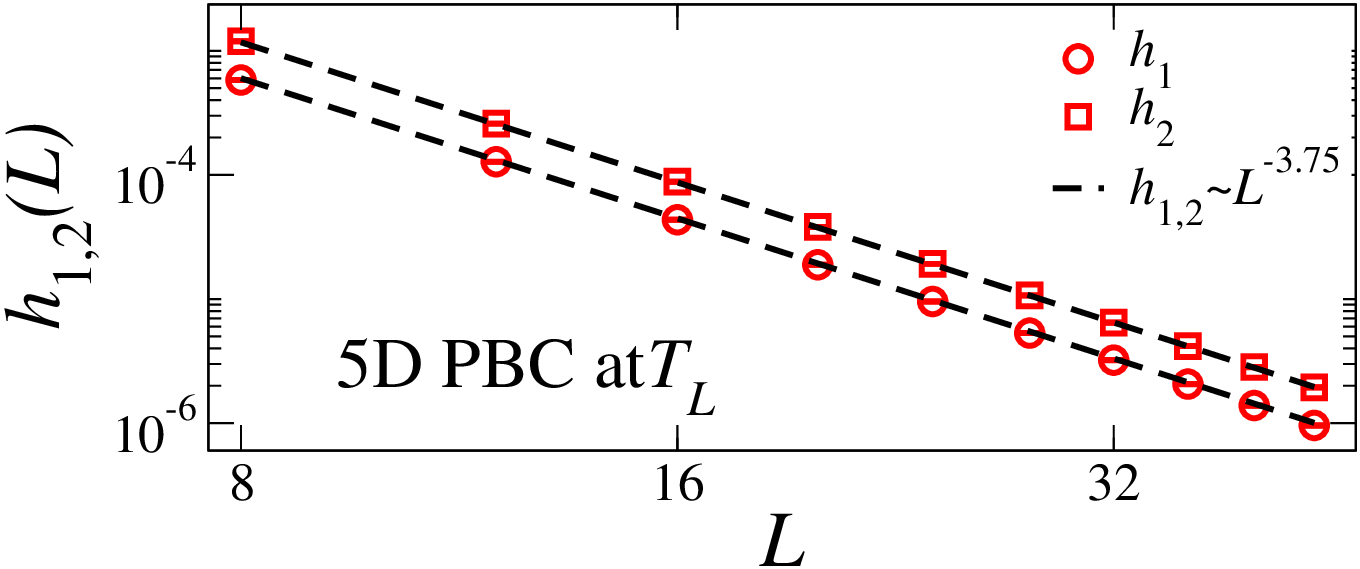}
  \includegraphics[width=0.8\columnwidth]{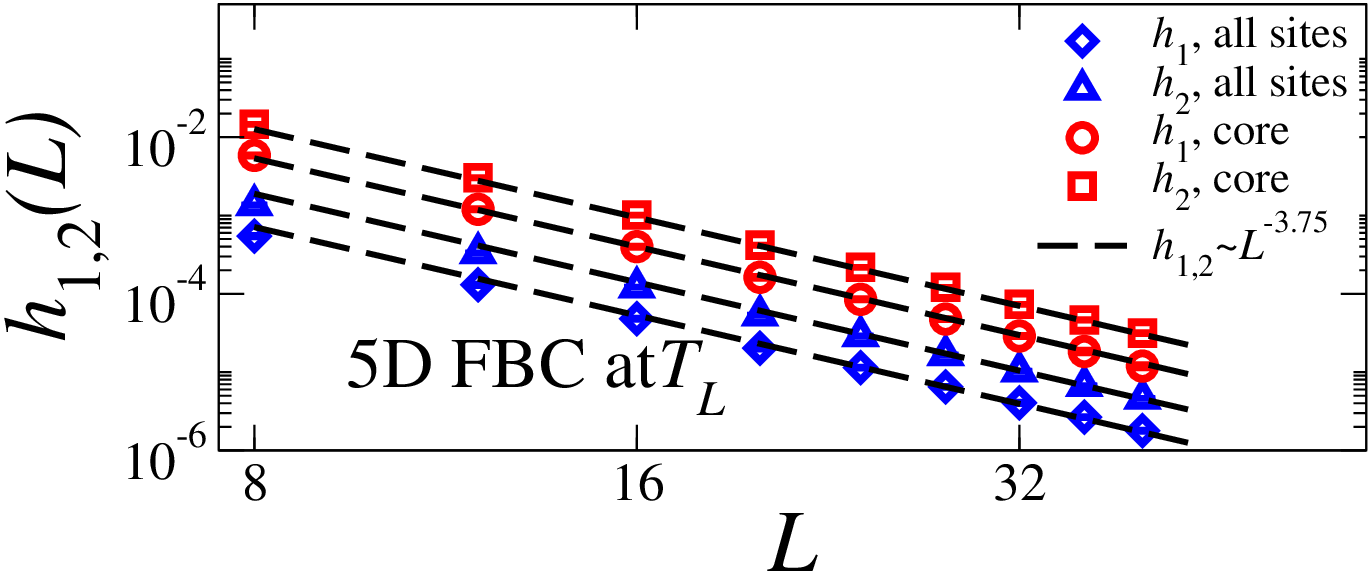}
  \caption{Analysis of the Lee-Yang zeroes (a) at criticality  and (b) at pseudocriticality for 5D PBC lattices ($L=8$ to 44 in steps of 4) as well as
(c) at pseudocriticality  for FBC lattices ($L=8$ to 32 in steps of 4).
In each case, the $Q$-FSS $h_{j}(L)\propto L^{-3.75}$ is confirmed, instead of the Gaussian  $h_j(L)\propto L^{-3}$. }
  \label{LY0_vs_L_forChi_5dPBC}
  \end{center}
\end{figure}

\section{Discussion}

Hyperscaling and finite-size scaling are intimately related. Both are universal below the upper critical dimension, but beyond that, the expectation until now has been that (a) hyperscaling is violated and (b) FSS becomes boundary-condition dependent.

Finite-size scaling below the upper critical dimension is very well understood as being controlled by the ratio of the correlation length to the actual length of a system, $\xi/L$. Since $\xi_\infty(t) \sim t^{-\nu}$, FSS there simply corresponds to the replacement $t \rightarrow L^{-1/\nu}$ in the scaling forms for the various functions. 
This prescription holds, and has been verified many times over, both at the infinite-volume critical point $T_c$ and at the pseudocritical point $T_L$ where thermodynamic functions peak.

Above the upper critical dimension, FSS with periodic boundary conditions is reasonably well understood through Eq.(\ref{BNPY}) and the dangerous irrelevant variable mechanism.
Eq.(\ref{BNPY2}) with $q \ne 1$ extends the mechanism to the correlation length.  
This renders the correlation length a power of the actual length of the system, $\xi \sim L^q$ with $q=d/d_c$.
FSS is implemented by the substitution $t \rightarrow L^{-q/\nu}$, and this is valid both at the critical and pseudocritical points.
This is called ``modified FSS'' or, for the purposes of this paper, ``$Q$-FSS''.
 
FSS with free boundaries has been less studied and is less understood above the upper critical dimension. 
One reason for the physical importance of this case is that systems with long-range interactions have reduced upper critical dimensions. If the interaction range is sufficiently long, such systems are experimentally accessible. In these cases, free boundary conditions represent the most realistic geometries.
This is the situation we have focused upon in this paper, using the five-dimensional (short-range) Ising model as a laboratory. 
The belief hitherto has been that $\xi \sim L$ so that a similar situation to that below $d=d_c$ prevails.
That is to say, ordinary FSS rather than $Q$-FSS was until now believed to be appropriate for FBC's above $d_c$.
Since the critical exponents above the critical dimension are those coming from Landau theory, we have termed the extension of ordinary FSS to $d>d_c$ as ``Gaussian''.

Here we have shown that, on the basis of extensive numerical calculations, there is in fact no support for Gaussian FSS at either criticality or pseudocriticality in the 5D Ising model.
We ascribe previous hints at Gaussian FSS as being due to very strong boundary effects -- the surface of the 5D FBC lattice is a 4D manifold, where Gaussian behaviour is indeed expected to leading order.
The new feature of our approach, which is absent in previous studies of this model, is that we have removed contributions of peripheral sites to truly probe the five-dimensionality of the FBC lattices.
We have also pointed out that previous analytical expectations for Gaussian FSS in the FBC case are not on solid ground.

Having dismissed Gaussian FSS in the FBC case, we present evidence in favour of $Q$-FSS at the pseudocritical point.
That this does not extend to the critical point (the focus of previous studies) is due to the fact that shifting is greater than rounding in this case. 
Shifting is well known to be boundary-condition dependent also below four dimensions, so this is not an infringement of universality.
Thus we arrive at our first main conclusion: $Q$-FSS at pseudocriticality in high-dimensional systems accurately describes FBC systems, as it does with PBC's and Gaussian FSS does not hold.

Physically,  ordinary FSS is viewed as being controlled by the ratio of the two length scales characterising the system, namely by $\xi/L$.
The physical interpretation of $Q$-FSS presented here is that the relevant ratio is rather of correlation volume to actual volume, namely $\xi^{d_c}/L^d$.
We also offer an analytical basis for our conclusions, using a Lee-Yang analysis. This brings us to our second main conclusion, namely that hyperscaling can be extended to beyond the  upper critical dimension in a universal manner. This is encapsulated in the new $Q$-version of Eq.(\ref{hyperhyperscaling}).
Since that equation relates $\alpha$ and $\nu$ to the new exponent $q$, and since the first two are universal, this indicates that $q$ is also a universal exponent and this is supported by our numerics for PBC's and FBC's.
In the case where $d \le d_c$, this exponent reverts to $q=1$.
The  new interpretation of FSS and hyperscaling subsumes the old one and holds for both PBC's and FBC's, above and below the upper critical dimension.
Therefore Eq.(\ref{hyperhyperscaling}) represents a universal version of hyperscaling, which is valid in all dimensions.

~ \\ 
~ \\
\noindent
{\bf{Acknowledgements:}}
We thank Christophe Chatelain and Yurij Holovatch for discussions.
R.K. is grateful for a Research Sabbatical Fellowship at Coventry University and for an invited  position at 
Nancy Universit{\'{e}}. 
This work is supported by the EU  FP7 IRSES Projects  269139 and 295302.

\bigskip
%

\end{document}